\documentclass[aps,twocolumn,groupedaddress,showpacs,floatfix]{revtex4}

\def\w{v}

\begin{document}
\bibliographystyle{apsrev}

\title{String-net condensation: A physical mechanism for topological phases}

\author{Michael A. Levin}
\author{Xiao-Gang Wen}
\homepage{http://dao.mit.edu/~wen}
\affiliation{Department of Physics, Massachusetts Institute of Technology,
Cambridge, Massachusetts 02139}
\date{April 2004}

\begin{abstract}
We show that quantum systems of extended objects naturally give rise to a
large class of exotic phases - namely topological phases. These phases occur
when the extended objects, called ``string-nets'', become highly fluctuating
and condense. We derive exactly soluble Hamiltonians for 2D local bosonic
models whose ground states are string-net condensed states. Those ground
states correspond to 2D parity invariant topological phases. These models
reveal the mathematical framework underlying topological phases: tensor
category theory. One of the Hamiltonians - a spin-1/2 system on the honeycomb
lattice - is a simple theoretical realization of a fault tolerant quantum
computer. The higher dimensional case also yields an interesting result: we
find that 3D string-net condensation naturally gives rise to both emergent
gauge bosons and emergent fermions. Thus, string-net condensation provides a
mechanism for unifying gauge bosons and fermions in 3 and higher dimensions.
\end{abstract}

\maketitle

\section{Introduction}

For many years, it was thought that Landau's theory of symmetry breaking
\cite{L3726} could describe essentially all phases and phase transitions. It
appeared that all continuous phase transitions were associated with a broken
symmetry. However, after the discovery of the fractional quantum Hall (FQH)
effect, it was realized that FQH states contain a new type of order 
- topological order - that is beyond the scope of Landau theory (for a review,
  see \Ref{Wtoprev}). Since then the study of topological phases in condensed
matter systems has been an active area of research. Topological phases have
been investigated in a variety of theoretical and experimental systems,
ranging from FQH systems \cite{WNtop,BWkmat2,R9002,FK9169}, quantum dimer
models \cite{RK8876,RC8933,MS0181,AFF0493} , quantum spin models
\cite{KL8795,WWZcsp,Wrig,RS9173,Wsrvb,SF0050,Wqoslpub,SP0258,BFG0212}, to
quantum computing \cite{K032,IFI0203}, or even superconducting states
\cite{Wtopcs,HOS0427}.  This work has revealed a host of interesting
theoretical phenomena and applications, including fractionalization, anyonic
quasiparticles, and fault tolerant quantum computation. Yet, a general theory
of topological phases is lacking.

One way to reveal the gaps in our understanding is to compare with Landau's
theory of symmetry breaking phases. Landau theory is based on (a) the physical
concepts of long range order, symmetry breaking, and order parameters, and (b)
the mathematical framework of group theory. These tools allow us to solve
three important problems in the study of ordered phases. First, they provide
low energy effective theories for general ordered phases: Ginzburg-Landau
field theories \cite{GL5064}. Second, they lead to a classification of
symmetry-breaking states. For example, we know that there are only 230
different crystal phases in three dimensions. Finally, they allow us to
determine the universal properties of the quasiparticle excitations (e.g.
whether they are gapped or gapless). In addition, Landau theory provides a
physical picture for the emergence of ordered phases - namely particle
condensation. 

Several components of Landau theory have been successfully reproduced in the
theory of topological phases. For example, the low energy behavior of
topological phases is relatively well understood on a formal level:
topological phases are gapped and are described by topological quantum field
theories (TQFT's).\cite{W8951} The problem of physically characterizing
topological phases has also been addressed. \Ref{Wtoprev} investigated the
``topological order'' (analogous to long range order) that occurs in
topological phases. The author showed that topological order is characterized
by robust ground state degeneracy, nontrivial particle statistics, and gapless
edge excitations.\cite{Wrig,WNtop,Wedgerev} These properties can be used to
partially classify topological phases. Finally, the quasiparticle excitations
of topological phases have been analyzed in particular cases. Unlike the
symmetry breaking case, the emergent particles in topologically ordered (or
more generally, quantum ordered) states include (deconfined) gauge
bosons\cite{BMK7793,FNN8035} as well as  fermions (in three dimensions)
\cite{LWsta,Wqoem} or anyons (in two dimensions) \cite{ASW8422}. Fermions and
anyons can emerge as collective excitations of purely bosonic models.

Yet, the theory of topological phases is still incomplete. The theory 
lacks two important components: a physical picture (analogous to particle 
condensation) that clarifies how topological phases emerge from microscopic 
degrees of freedom, and a mathematical framework (analogous to group theory) 
for characterizing and classifying these phases.

In this paper, we address these two issues for a large class of topological
phases which we call ``doubled'' topological phases. On a formal level, 
``doubled'' topological phases are phases that are described by a sum of two 
TQFT's with opposite chiralities. Physically, they are characterized by parity 
and time reversal invariance. Examples include all discrete lattice gauge 
theories, and all doubled Chern-Simons theories. It is unclear to what extent 
our results generalize to chiral topological phases - such as in the FQH 
effect.

\begin{figure}[tb]
\centerline{
\includegraphics[width=2.5in]{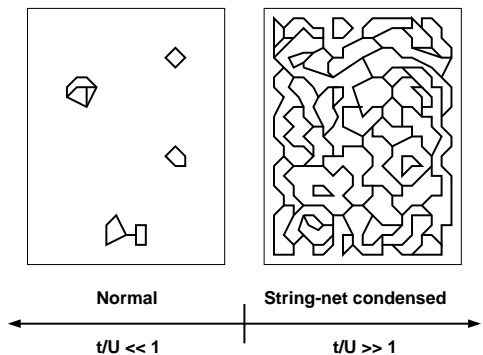}
}
\caption{
A schematic phase diagram for the generic string-net Hamiltonian (\ref{genh}).
When $t/U$ (the ratio of the kinetic energy to the string tension) is small 
the system is in the normal phase. The ground state is essentially the vacuum 
with a few small string-nets. When $t/U$ is large the string-nets condense and 
large fluctuating string-nets fill all of space. We expect a phase transition 
between the two states at some $t/U$ of order unity. We have omitted string 
labels and orientations for the sake of clarity. 
}
\label{strphase}
\end{figure}

We first address the problem of the physical picture for doubled topological
phases. We argue that in these phases, local energetic constraints cause the
microscopic degrees of freedom to organize into effective extended objects
called ``string-nets''. At low energies, the microscopic Hamiltonian 
effectively describes the dynamics of these extended objects. If the kinetic 
energy of the string-nets dominates the string-net tension, the string-nets 
``condense'': large string-nets with a typical size on the same order as the 
system size fill all of space (see Fig. \ref{strphase}). The result is a 
doubled topological phase. Thus, just as traditional ordered phases arise via 
particle condensation, topological phases originate from ``string-net 
condensation.''

This physical picture naturally leads to a solution to the second problem -
that of finding a mathematical framework for classifying and characterizing
doubled topological phases. We show that each topological phase is associated
with a mathematical object known as a ``tensor category.'' \cite{Kas95} Here, 
we think of a tensor category as a $6$ index object $F^{ijk}_{lmn}$ which 
satisfies certain algebraic equations \Eq{pent}. The mathematical object 
$F^{ijk}_{lmn}$ characterizes different topological phases and determines the 
universal properties of the quasiparticle excitations (e.g.  statistics) just 
as the symmetry group does in Landau theory. We feel that the mathematical 
framework of tensor categories, together with the physical picture of 
string-net condensation provides a general theory of (doubled) topological 
phases. 

Our approach has the additional advantage of providing exactly soluble
Hamiltonians and ground state wave functions for each of these phases.  Those
exactly soluble Hamiltonians describe local bosonic models (or spin models).
They realize all discrete gauge theories (in any dimension) and all doubled
Chern-Simons theories (in $(2+1)$ dimensions). One of the Hamiltonians - a
spin-1/2 model on the honeycomb lattice - is a simple theoretical realization
of a fault tolerant quantum computer \cite{FLZ0205}. The higher dimensional
models also yield an interesting result: we find that $(3+1)D$ string-net
condensation naturally gives rise to both emerging gauge bosons and emerging
fermions. Thus, string-net condensation provides a mechanism for unifying
gauge bosons and fermions in $(3+1)$ and higher dimensions.

We feel that this constructive approach is one of the most important features
of this paper. Indeed, in the mathematical community it is well known that
topological field theory, tensor category theory and knot theory are all
intimately related \cite{Tur94,W8929,W9085}. Thus it is not surprising that
topological phases are closely connected to tensor categories and string-nets.
The contribution of this paper is our demonstration that these elegant
mathematical relations have a concrete realization in condensed matter
systems.

The paper is organized as follows. In sections II and III, we introduce the
string-net picture, first in the case of deconfined gauge theories, and then 
in the general case. We argue that all doubled topological phases are 
described by string-net condensation. 

The rest of the paper is devoted to developing a theory of string-net
condensation. In section IV, we consider the case of $(2+1)$ dimensions. 
In parts A and B, we construct string-net wave functions and Hamiltonians 
for each $(2+1)D$ string-net condensed phase. Then, in part C, we use this
mathematical framework to calculate the universal properties of the 
quasiparticle excitations in each phase. In section V, we discuss the 
generalization to $3$ and higher dimensions. In the last section, we present 
several examples of string-net condensed states - including a spin-1/2 model
theoretically capable of fault tolerant quantum computation. The main 
mathematical calculations can be found in the appendix.   

\section{String-nets and gauge theories}

In this section, we introduce the string-net picture in the context of
gauge theory \cite{KS7595,BMK7793,Walight}. We point out that all 
deconfined gauge theories can be 
understood as string-net condensates where the strings are essentially
electric flux lines. We hope that this result provides intuition 
for (and motivates) the string-net picture in the general case.

\begin{figure}[tb]
\centerline{
\includegraphics[width=2.5in]{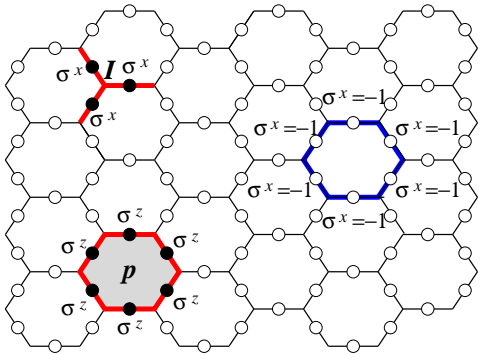}
}
\caption{
The constraint term $\prod_{\text{legs of }\v I} \si^x_{\v i}$ and magnetic 
term $\prod_{\text{edges of }\v p} \si^z_{\v j}$ in $Z_2$ lattice gauge theory.
In the dual picture, we regard the links with $\si^x = -1$ as being occupied by
a string, and the links with $\si^x = +1$ as being unoccupied. The constraint
term then requires the strings to be closed - as shown on the right.
}
\label{Z2gaugH}
\end{figure}

We begin with the simplest gauge theory - $Z_2$ lattice gauge theory 
\cite{W7159}. The Hamiltonian is
\begin{equation}
\label{Z2gaugHam}
 H_{Z_2}= 
-U\sum_{\v i}  \si^x_{\v i}
+t\sum_{\v p}  \prod_{\text{edges of }\v p} \si^z_{\v j}
\end{equation}
where $\si^{x,y,z}$ are the Pauli matrices, and $\v I$, $\v i$, $\v p$ label 
the sites, links, and plaquettes of the lattice. The Hilbert space is
formed by states satisfying
\begin{equation}
\label{Z2const}
 \prod_{\text{legs of }\v I} \si^x_{\v i} |\Phi\>=|\Phi\>,
\end{equation}
for every site $\v I$. For simplicity we will restrict our discussion to 
trivalent lattices such as the honeycomb lattice (see Fig. \ref{Z2gaugH}).

It is well known that $Z_2$ lattice gauge theory is dual to the Ising model in
$(2+1)$ dimensions \cite{K7959}. What is less well known is that there is a
more general dual description of $Z_2$ gauge theory that exists in any number
of dimensions \cite{ItzSZ88}. To obtain this dual picture, we view links 
with $\si^x=-1$ as being occupied by a string and links with $\si^x=+1$ as 
being unoccupied. The constraint \Eq{Z2const} then implies that only 
closed strings are allowed in the Hilbert space (Fig. \ref{Z2gaugH}). 
  
\begin{figure}[tb]
\centerline{
\includegraphics[width=3in]{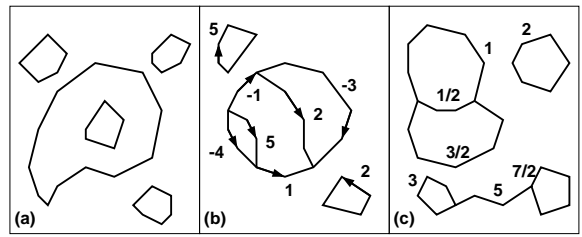}
}
\caption{
Typical string-net configurations in the dual formulation of (a) $Z_2$, 
(b) $U(1)$, and (c) $SU(2)$ gauge theory. In the case of (a) $Z_2$ gauge 
theory, the string-net configurations consist of closed (non-intersecting) 
loops. In (b) $U(1)$ gauge theory, the string-nets are oriented
graphs with edges labeled by integers. The string-nets obey the branching rules
$E_1 + E_2 + E_3 = 0$ for any three edges meeting at a point. In the case of 
(c) $SU(2)$ gauge theory, the string-nets consist of (unoriented) graphs with 
edges labeled by half-integers $1/2,1,3/2,...$. The branching rules are given 
by the triangle inequality:  $\{E_1,E_2,E_3\}$ are allowed to meet at a point 
if and only if $E_1 \leq E_2 + E_3$, $E_2 \leq E_3 + E_1$, 
$E_3 \leq E_1 + E_2$, and $E_1+E_2+E_3$ is an integer.
}
\label{strgauge}
\end{figure}

In this way, $Z_2$ gauge theory can be reformulated as a closed string theory,
and the Hamiltonian can be viewed as a closed string Hamiltonian. The electric
and magnetic energy terms have a simple interpretation in this dual picture:
the ``electric energy'' $-U\sum_{\v i}  \si^x_{\v i}$ is a string tension
while the ``magnetic energy'' $t\sum_{\v p}  \prod_{\text{edges of }\v p}
\si^z_{\v j}$ is a string kinetic energy. The physical picture for the
confining and deconfined phases is also clear. The confining phase corresponds
to a large electric energy and hence a large string tension $U \gg t$.  The
ground state is therefore the vacuum configuration with a few small strings.
The deconfined phase corresponds to a large magnetic energy and hence a large
kinetic energy. The ground state is thus a superposition of many large string
configurations. In other words, the deconfined phase of $Z_2$ gauge theory is
a quantum liquid of large strings - a ``string condensate.'' (Fig.
\ref{strgauge}a).

A similar, but more complicated, picture exists for other deconfined
gauge theories. The next layer of complexity is revealed when we consider
other Abelian theories, such as $U(1)$ gauge theory. As in the case of
$Z_2$, $U(1)$ lattice gauge theory can be reformulated as a theory of 
electric flux lines. However, unlike $Z_2$, there is more then one type of 
flux line. The electric flux on a link can take any integral value in $U(1)$
lattice gauge theory. Therefore, the electric flux lines need to be labeled
with integers to indicate the amount of flux carried by the line. In
addition, the flux lines need to be oriented to indicate the direction of 
the flux. The final point is that the flux lines don't necessarily form closed 
loops. It is possible for three flux lines $E_1, E_2, E_3$ to meet at a 
point, as long as Gauss' law is obeyed: $E_1 + E_2 + E_3 = 0$. Thus, the 
dual formulation of $U(1)$ gauge theory involves not strings, but 
more general objects: networks of strings (or ``string-nets''). The 
strings in a string-net are labeled, oriented, and obey branching rules,
given by Gauss' law (Fig. \ref{strgauge}b). 

This ``string-net'' picture exists for general gauge theories. In the 
general case, the strings (electric flux lines) are labeled by 
representations of the gauge group. The branching rules (Gauss' law) 
require that if three strings $E_1, E_2, E_3$ meet at a point, then the 
product of the representations $E_1 \otimes E_2 \otimes E_3$ must contain the 
trivial representation. (For example, in the case of $SU(2)$, the strings
are labeled by half-integers $E = 1/2, 1, 3/2, ...$, and the branching  
rules are given by the triangle inequality: $\{E_1,E_2,E_3\}$ are 
allowed to meet at a point if and only if $E_1 \leq E_2 + E_3$, $E_2 \leq 
E_3 + E_1$, $E_3 \leq E_1 + E_2$ and $E_1+E_2+E_3$ is an integer 
(Fig. \ref{strgauge}c)) \cite{KS7595}. These string-nets provide a general dual
formulation of gauge theory. As in the case of $Z_2$, the deconfined phase of 
the gauge theory always corresponds to highly fluctuating string-nets -- a 
string-net condensate. 

\section{General string-net picture}
Given the large scope of gauge theory, it is natural to wonder if 
string-nets can describe more general topological phases.
In this section we will discuss this more general string-net picture.
(Actually, we will not discuss the \emph{most} general string-net picture. 
We will focus on a special case for the sake of simplicity. See
Appendix \ref{genstr} for a discussion of the most general picture).

\begin{figure}[tb]
\centerline{
\includegraphics[width=0.5in]{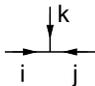}
}
\caption{
The orientation convention for the branching rules.
}
\label{strBrO}
\end{figure}

\begin{figure}[tb]
\centerline{
\includegraphics[width=1.2in]{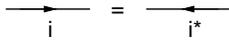}
}
\caption{
$i$ and $i^*$ label strings with opposite orientations.
}
\label{strOrt}
\end{figure}

We begin with a more detailed definition of ``string-nets.'' As the name 
suggests, string-nets are networks of strings. We will focus on trivalent 
networks where each node or branch point is attached to exactly $3$ strings. 
The strings in a string-net are oriented and come in various ``types.'' Only 
certain combinations of string types are allowed to meet at a node or branch 
point. To specify a particular string-net model, one needs to provide the
following data:

\begin{enumerate}
\item
\textbf{String types}: The number of different string types $N$. For 
simplicity, we will label the different string types with the integers
$i = 1, ..., N$.  
\item
\textbf{Branching rules}: The set of all triplets of string-types $\{\{i,j,k\}
... \}$ that are allowed to meet at a point. (See Fig. \ref{strBrO}).
\item
\textbf{String orientations}: The dual string type $i^*$ associated with each
string type $i$. The duality must satisfy $(i^*)^* = i$. The type-$i^*$ string
corresponds to the type-$i$ string with the opposite orientation. If 
$i = i^*$, then the string is unoriented (See Fig. \ref{strOrt}).   
\end{enumerate}

This data describes the detailed structure of the string-nets. The Hilbert
space of the string-net model is then defined in the natural way. The
states in the Hilbert space are simply linear superpositions of different 
spatial configurations of string-nets. 

Once the Hilbert space has been specified, we can imagine writing down
a string-net Hamiltonian. The string-net Hamiltonian can be any local operator
which acts on quantum string-net states. A typical Hamiltonian is a sum of
potential and kinetic energy pieces: 
\begin{equation}
H = U H_{U} + t H_{t}
\label{genh}
\end{equation} 
The kinetic energy $H_{t}$ gives dynamics to the string-nets, while the
potential energy $H_{U}$ is typically some kind of string tension. When $U >>
t$, the string tension dominates and we expect the ground state to be the
vacuum state with a few small string-nets.  On the other hand, when $t >> U$,
the kinetic energy dominates, and we expect the ground state to consist of
many large fluctuating string-nets.  We expect that there is a quantum phase
transition between the two states at some $t/U$ on the order of unity.  (See
Fig. \ref{strphase}). Because of the analogy with particle condensation, we
say that the large $t$, highly fluctuating string-net phase is ``string-net
condensed.''

This notion of string-net condensation provides a natural physical mechanism 
for the emergence of topological phases in real condensed matter systems. Local
energetic constraints can cause the microscopic degrees of freedom to organize 
into effective extended objects or string-nets. If the kinetic energy of these 
string-nets is large, then they can condense giving rise to a topological 
phase. The type of topological phase is determined by the structure of the 
string-nets, and the form of string-net condensation.

But how general is this picture? In the previous section, we pointed out
that all deconfined gauge theories can be viewed as string-net condensates.
In fact, mathematical results suggest that the string-net picture is even
more general. In $(2+1)$ dimensions, all so-called ``doubled'' topological 
phases can be described by string-net condensation (provided that we
generalize the string-net picture as in Appendix \ref{genstr}). \cite{Tur94} 
Physically, this means that the string-net picture can be applied to 
essentially all parity and time reversal invariant topological phases in 
$(2+1)$ dimensions. Examples include all discrete gauge theories, and all 
doubled Chern-Simons theories. The situation for dimension $d > 2$ is less well
understood. However, we know that string-net condensation quite generally 
describes all lattice gauge theories with or without emergent Fermi statistics.

\section{String-net condensation in $(2+1)$ dimensions}

\subsection{Fixed-point wave functions}
In this section, we attempt to capture the universal features of  string-net
condensed phases in $(2+1)$ dimensions. Our approach, inspired by
\Ref{KS7595,W8929,W9085,F0160,FNS0311,FNS0320}, is based on the string-net
wave function. We construct a special ``fixed-point'' wave function for each
string-net condensed phase. We believe that these ``fixed-point'' wave
functions capture the universal properties of the corresponding phases. Each
``fixed-point'' wave function is associated with a six index object
$F^{ijk}_{lmn}$ that satisfies certain algebraic equations $(\ref{pent})$. In
this way, we derive a one-to-one correspondence between doubled topological
phases and tensor categories $F^{ijk}_{lmn}$. We would like to mention that a
related result on the classification of $(2+1)D$ topological quantum field
theories was obtained independently in the mathematical community.
\cite{Tur94}

Let us try to visualize the wave function of a string-net condensed state. 
Though we haven't defined string-net condensation rigorously, we expect that a 
string-net condensed state is a superposition of many different string-net 
configurations. Each string-net configuration has a size typically on the same
order as the system size. The large size of the string-nets implies that a 
string-net condensed wave function has a non-trivial long distance structure. 
It is this long distance structure that distinguishes the condensed state from 
the ``normal'' state. 

\begin{figure}[tb]
\centerline{
\includegraphics[width=1.5in]{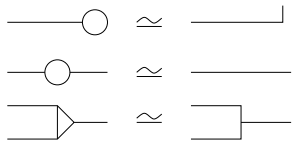}
}
\caption{
Three pairs of string-net configurations that differ only in their
short distance structure. We expect string-net wave functions in the
same quantum phase to only differ by these short distance details.
}
\label{lngftr}
\end{figure}

In general, we expect that the universal features of a string-net condensed 
phase are contained in the long distance character of the wave functions. 
Imagine comparing two different string-net condensed states that belong to
the same quantum phase. The two states will have different wave functions.
However, by the standard RG reasoning, we expect that the two wave functions 
will look the same at long distances. That is, the two wave functions will only
differ in short distance details - like those shown in Fig. \ref{lngftr}.

\begin{figure}[tb]
\centerline{
\includegraphics[width=2.in]{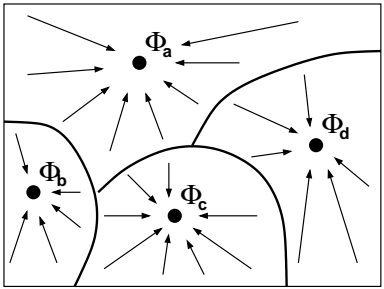}
}
\caption{
A schematic RG flow diagram for a string-net model with $4$ string-net
condensed phases $a$, $b$, $c$, and $d$. All the states in each phase flow
to fixed-points in the long distance limit. The corresponding fixed-point wave 
functions $\Phi_{a}$, $\Phi_{b}$, $\Phi_{c}$, and $\Phi_{d}$ capture the 
universal long distance features of the associated quantum phases. Our
ansatz is that the fixed-point wave functions $\Phi$ are described by 
local constraints of the form (\ref{topinv}-\ref{fusion}).  
}
\label{fixpt}
\end{figure}

Continuing with this line of thought, we imagine performing an RG analysis on
ground state functions. All the states in a string-net condensed phase should 
flow to some special ``fixed-point'' state. We expect that the wave function of
this state captures the universal long distance features of the whole quantum 
phase. (See Fig. \ref{fixpt}).

In the following, we will construct these special fixed-point wave functions.
Suppose $\Phi$ is some fixed-point wave function. We know that $\Phi$ is the
ground state of some fixed-point Hamiltonian $H$. Based on our experience with
gauge theories, we expect that $H$ is free. That is, $H$ is a sum of local 
string kinetic energy terms with no string tension terms:
\begin{displaymath}
H = tH_t = t\sum_i H_{t,i}
\end{displaymath}
In particular, $H$ is unfrustrated, and the ground state wave function 
minimizes the expectation values of all the kinetic energy terms $\{H_{ti}\}$ 
simultaneously. Minimizing the expectation value of an individual kinetic 
energy term $H_{t,i}$ is equivalent to imposing a local constraint on the 
ground state wave function, namely $H_{t,i}|\Phi\> = E_{i} |\Phi\>$ (where
$E_{i}$ is the smallest eigenvalue of $H_{t,i}$). We conclude that 
\emph{the wave function $\Phi$ can be specified uniquely by local constraint 
equations}. The local constraints are linear relations between several 
string-net amplitudes $\Phi(X_1),\Phi(X_2),\Phi(X_3)...$ where the 
configurations $X_1,X_2,X_3...$ only differ by local transformations. 

To derive these local constraints from first principles is difficult, so we 
will use a more heuristic approach. We will first guess the form of the local 
constraints (\ie guess the form of the fixed-point wave function). Then, in the
next section, we will construct the fixed-point Hamiltonian 
and show that its ground state wave function does indeed satisfy these local 
relations. Our ansatz is that the local constraints can be put in the following
graphical form:
\begin{align}
 \Phi
\bpm \includegraphics[height=0.3in]{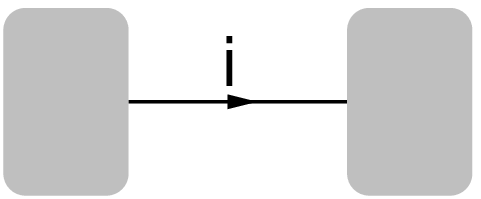} \epm  =&
\Phi 
\bpm \includegraphics[height=0.3in]{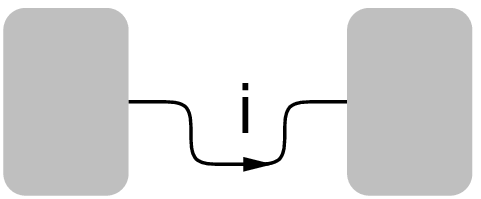} \epm
\label{topinv}
\\
 \Phi
\bpm \includegraphics[height=0.3in]{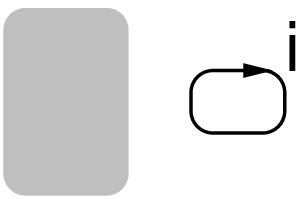} \epm  =&
d_i\Phi 
\bpm \includegraphics[height=0.3in]{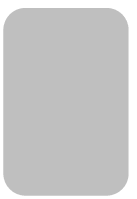} \epm
\label{clsdst}
\\
 \Phi
\bpm \includegraphics[height=0.3in]{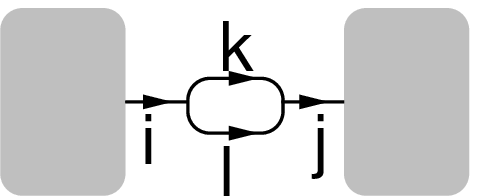} \epm  =&
\delta_{ij}
\Phi 
\bpm \includegraphics[height=0.3in]{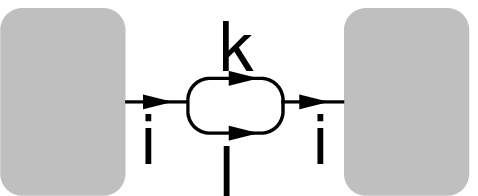} \epm
\label{bubble}\\
 \Phi
\bpm \includegraphics[height=0.3in]{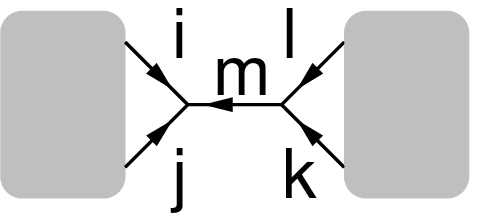} \epm  =&
\sum_{n} 
F^{ijm}_{kln}
\Phi 
\bpm \includegraphics[height=0.28in]{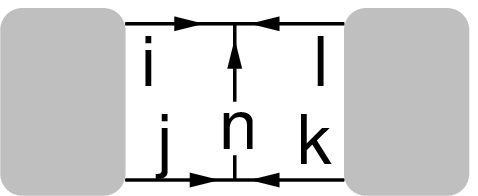} \epm
\label{fusion} 
\end{align}
Here, $i$, $j$, $k$ etc. are arbitrary string types and the shaded regions
represent arbitrary string-net configurations. The $d_i$ are complex numbers.
The $6$ index symbol $F^{ijm}_{kln}$ is a complex numerical constant that
depends on $6$ string types $i$, $j$, $m$, $k$, $l$, and $n$. If one or more
of the branchings $\{i,j,m\},\{k,l,m^*\}, \{i,n,l\},\{j,k,n^*\}$ is illegal,
the value of the symbol  $F^{ijm}_{kln}$ is unphysical. However, for
simplicity, we will set $F^{ijm}_{kln} = 0$ in this case. 

The local rules (\ref{topinv}-\ref{fusion}) are written using a new notational 
convention. According to this convention, the indices $i,j,k$ etc., can take on
the value $i = 0$ in addition to the $N$ physical string types $i = 1, ... N$. 
We think of the $i=0$ string as the ``empty string'' or ``null string.'' It 
represents empty space - the vacuum. Thus, we can convert labeled string-nets 
to our old convention by simply erasing all the $i = 0$ strings. The branching 
rules and dualities associated with $i=0$ are defined in the obvious way:
$0^* = 0$, and $\{i,j,0\}$ is allowed if and only if $i = j^*$. Our convention 
serves two purposes: it simplifies notation (each equation in (\ref{topinv}-
\ref{fusion}) represents several equations with the old convention), and it 
reveals the mathematical framework underlying string-net condensation. 

We now briefly motivate these rules.
The first rule (\ref{topinv}) constrains the wave function $\Phi$ to be 
topologically invariant. It requires the quantum mechanical amplitude for 
a string-net configuration to only depend on the topology of the 
configuration: two configurations that can be continuously deformed into 
one another must have the same amplitude. The motivation for this constraint 
is our expectation that topological string-net phases have topologically 
invariant fixed-points. 

The second rule (\ref{clsdst}) is motivated by the fundamental property of
RG fixed-points: scale invariance. The wave function $\Phi$ should look
the same at all distance scales. Since a closed string
disappears at length scales larger then the string size, the 
amplitude of an arbitrary string-net configuration with a closed string
should be proportional to the amplitude of the string-net configuration
alone. 

The third rule (\ref{bubble}) is similar. Since a ``bubble'' is irrelevant at 
long length scales, we expect  
\begin{align*}
 \Phi
\bpm \includegraphics[height=0.3in]{XikljO.eps} \epm  &\propto
 \Phi
\bpm \includegraphics[height=0.3in]{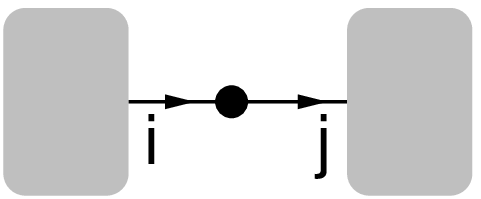} \epm
\end{align*}
But if $i \neq j$, the configuration 
$\includegraphics[height=0.3in]{XijO.eps}$ is not allowed: 
$\Phi\bpm \includegraphics[height=0.3in]{XijO.eps} \epm = 0$. We conclude that
the amplitude for the bubble configuration vanishes when $i \neq j$ 
(\ref{bubble}).

The last rule is less well-motivated. The main point is that the first three 
rules are not complete: another constraint is needed to specify the ground 
state wave function uniquely. The last rule (\ref{fusion}) is the simplest
local constraint with this property. An alternative motivation for this
rule is the fusion algebra in conformal field theory.\cite{MS8977}

The local rules (\ref{topinv}-\ref{fusion}) uniquely specify the fixed-point 
wave function $\Phi$. The universal features of the string-net condensed state
are captured by these rules. Equivalently, they are captured by the six index 
object $F^{ijm}_{kln}$, and the numbers $d_i$. 

However, not every choice of $(F^{ijm}_{kln}$, $d_i)$ corresponds to a 
string-net condensed phase. In fact, a generic choice of $(F^{ijm}_{kln}$,
$d_i)$ will lead to constraints (\ref{topinv}-\ref{fusion}) that are not 
self-consistent. The only $(F^{ijm}_{kln}$, $d_i)$ that give rise to 
self-consistent rules and a well-defined wave function $\Phi$ are (up to a 
trivial rescaling) those that satisfy 
\begin{eqnarray}
F^{ijk}_{j^*i^*0} &=& \frac{\w_{k}}{\w_{i}\w_{j}} \del_{ijk} \nonumber \\ 
F^{ijm}_{kln} = F^{lkm^*}_{jin} &=& F^{jim}_{lkn^*} = F^{imj}_{k^*nl}
\frac{\w_{m}\w_{n}}{\w_{j}\w_{l}}
\nonumber \\
\sum_{n=0}^N 
F^{mlq}_{kp^*n} F^{jip}_{mns^*} F^{js^*n}_{lkr^*}
&=&
F^{jip}_{q^*kr^*} F^{riq^*}_{mls^*} 
\label{pent}
\end{eqnarray}
where $\w_{i} = \w_{i^*} = \sqrt{d_{i}}$ (and $\w_0 = 1$). 
(See appendix \ref{scs}). Here, we have introduced a new object $\del_{ijk}$ 
defined by the branching rules:
\begin{equation} 
\label{del}
\del_{ijk}=\left\{
\bmm
1, & \text{if $\{i,j,k\}$ is allowed,} \\
0, & \text{otherwise.}
\emm
\right. 
\end{equation} 

There is a one-to-one correspondence between $(2+1)D$ string-net condensed 
phases and solutions of \Eq{pent}. These solutions correspond to mathematical
objects known as tensor categories. \cite{Kas95} Tensor category theory is the 
fundamental mathematical framework for string-net condensation, just as group 
theory is for particle condensation. We have just shown that it gives a 
complete classification of $(2+1)D$ string-net condensed phases (or 
equivalently doubled topological phases): each phase is associated with a 
different solution to (\ref{pent}). We will show later that it also provides a 
convenient framework for deriving the physical properties of quasiparticles. 

It is highly non-trivial to find solutions of (\ref{pent}). 
However, it turns out each group $G$ provides a solution. The solution is
obtained by (a) letting the string-type index $i$ run over the irreducible 
representations of the group, (b) letting the numbers $d_i$ be the dimensions 
of the representations and (c) letting the $6$ index object $F^{ijm}_{kln}$ 
be the $6j$ symbol of the group. The low energy effective theory of the 
corresponding string-net condensed state turns out to be a deconfined gauge 
theory with gauge group $G$. Another class of solutions can be obtained from 
$6j$ symbols of quantum groups. It turns out that in these cases, the low 
energy effective theories of the corresponding string-net condensed states are 
doubled Chern-Simons gauge theories. These two classes of solutions are
not necessarily exhaustive: Eq. (\ref{pent}) may have solutions other then 
gauge theories or Chern-Simons theories. Nevertheless, it is clear that
gauge bosons and gauge groups emerge from string-net condensation in a very 
natural way.

In fact, string-net condensation provides a new perspective on gauge theory.
Traditionally, we think of gauge theories geometrically. The gauge field
$A_\mu$ is analogous to an affine connection, and the field strength 
$F_{\mu\nu}$ is essentially a curvature tensor. From this point of view, gauge 
theory describes the dynamics of certain geometric objects (e.g. fiber 
bundles). The gauge group determines the structure of these objects and 
is introduced by hand as part of the basic definition of the theory. 
In contrast, according to the string-net condensation picture, the
geometrical character of gauge theory is not fundamental. Gauge theories are 
fundamentally theories of extended objects. The gauge group and the 
geometrical gauge structure emerge dynamically at low energies and long 
distances. A string-net system ``chooses'' a particular gauge group, depending
on the coupling constants in the underlying Hamiltonian: these parameters
determine a string-net condensed phase which in turn determines a solution to
(\ref{pent}). The nature of this solution determines the gauge group.

One advantage of this alternative picture is that it unifies two seemingly 
unrelated phenomena: gauge interactions and Fermi statistics. Indeed, as we 
will show in section V, string-net condensation naturally gives rise to both 
gauge interactions and Fermi statistics (or fractional statistics in $(2+1)D$).
In addition, these structures always appear together. \cite{LWsta}
 
\subsection{Fixed-point Hamiltonians}

In this section, we construct exactly soluble lattice spin Hamiltonians with
the fixed-point wave functions $\Phi$ as ground states. These Hamiltonians
provide an explicit realization of all $(2+1)D$ string-net condensates and
therefore all $(2+1)D$ doubled topological phases (provided that we generalize
these models as discussed in Appendix \ref{genstr}). In the next section, we 
will use them to calculate the physical properties of the quasiparticle 
excitations.

For every $(F^{ijm}_{kln}, d_i)$ satisfying the self-consistency conditions
(\ref{pent}) and the unitarity condition (\ref{unit}), we can construct an 
exactly soluble Hamiltonian. Let us first describe the Hilbert space of the 
exactly soluble model. The model is a spin system on a (2D) honeycomb lattice, 
with a spin located on each link of the lattice. Each ``spin'' can be in 
$N+1$ different states labeled by $i=0,1,...,N$. We assign each link an 
arbitrary orientation. When a spin is in state $i$, we think of the link 
as being occupied by a type-$i$ string oriented in the appropriate direction. 
We think of the type-$0$ string or null string as the vacuum (\ie no string 
on the link).

\begin{figure}[tb]
\centerline{
\includegraphics[width=3in]{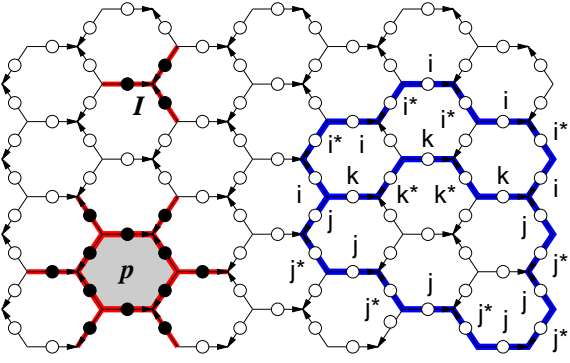}
}
\caption{
A picture of the lattice spin model (\ref{HPi}). The electric charge operator 
$Q_{\v I}$ acts on the three spins adjacent to the vertex $\v I$, while the 
magnetic energy operator $B_{\v p}$ acts on the $12$ spins adjacent to the 
hexagonal plaquette $\v p$. The term $Q_{\v I}$ constrains the string-nets to 
obey the branching rules, while $B_{\v p}$ provides dynamics. A typical state 
satisfying the low-energy constraints is shown on the right. The empty
links have spins in the $i = 0$ state.  
}
\label{SSpin}
\end{figure}

The exactly soluble Hamiltonian for our model is given by
\begin{equation}
\label{HPi}
 H= - \sum_{\v I} Q_{\v I}-\sum_{\v p} B_{\v p}, \ \ \ \ \  B_{\v p}= \sum_{s=0}^N a_s B_{\v p}^s
\end{equation}
where the sums run over vertices $\v I$ and plaquettes $\v p$ of the honeycomb 
lattice. The coefficients $a_s$ satisfy $a_{s^*} = a_s^*$ but are otherwise
arbitrary.

Let us explain the terms in \Eq{HPi}. We think of the first term
$Q_{\v I}$ as an electric charge operator. It measures the ``electric 
charge'' at site $\v I$, and favors states with no charge. It acts on
the $3$ spins adjacent to the site $\v I$:
\begin{equation}
Q_{\v I}
\left | \bmm\includegraphics[height=0.3in]{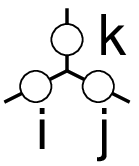}\emm \right\>
=
\del_{ijk}
\left | \bmm\includegraphics[height=0.3in]{starD.eps}\emm \right\>
\end{equation}
where $\del_{ijk}$ is the branching rule symbol (\ref{del}). Clearly,
this term constrains the strings to obey the branching rules described by
$\del_{ijk}$. With this constraint the low energy Hilbert
space is essentially the set of all allowed string-net configurations on
a honeycomb lattice. (See Fig. \ref{SSpin}).

We think of the second term $B_{\v p}$ as a magnetic flux operator. It
measures the ``magnetic flux'' though the plaquette $\v p$ (or more precisely,
the cosine of the magnetic flux) and favors states with no flux. This term 
provides dynamics for the string-net configurations. 

The magnetic flux operator $B_{\v p}$ is a linear combination of $(N+1)$ terms
$B_{\v p}^s$, $s=0,1,...,N$. Each $B_{\v p}^s$ is an operator that acts on the
$12$ links that are adjacent to vertices of the hexagon $\v p$. (See Fig.
\ref{SSpin}).  Thus, the $B_{\v p}^s$ are essentially $(N+1)^{12} \times
(N+1)^{12}$ matrices.  However, the action of $B_{\v p}^s$ does not change the
spin states on the $6$ outer links of $\v p$. Therefore the $B_{\v p}^s$ can
be block diagonalized into $(N+1)^6$ blocks, each of dimension $(N+1)^6\times
(N+1)^6$. Let $B^{s,g'h'i'j'k'l'}_{\v p,ghijkl}(abcdef)$, with $a,b,c... 
=0,1,...,N$, denote the matrix elements of these $(N+1)^6$ matrices:

\begin{align}
\label{BB}
&  B^s_{\v p} 
\left | \bmm\includegraphics[height=0.6in]{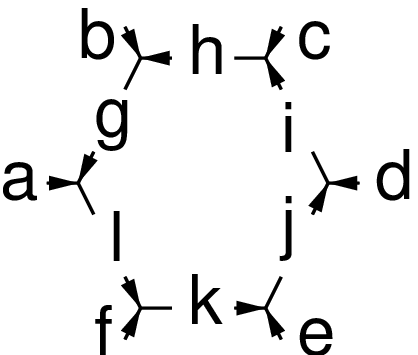}\emm \right\>
\nonumber\\
=&  \sum_{m,...,r} B_{\v p,ghijkl}^{s,g'h'i'j'k'l'}(abcdef)
\left | \bmm\includegraphics[height=0.6in]{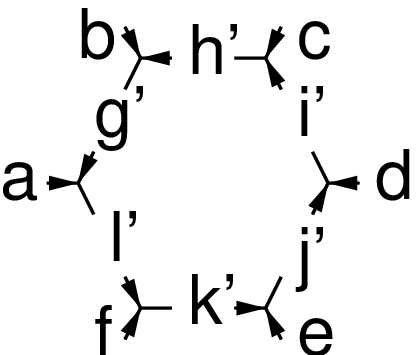}\emm \right\>
\end{align}
Then the operators $B_{\v p}^s$ are defined by
\begin{align}
\label{B6F}
&B_{\v p,ghijkl}^{s,g'h'i'j'k'l'}(abcdef)
 \nonumber \\
=&\;
F^{al^*g}_{s^*g'l'^*}
F^{bg^*h}_{s^*h'g'^*}
F^{ch^*i}_{s^*i'h'^*}
F^{di^*j}_{s^*j'i'^*}
F^{ej^*k}_{s^*k'j'^*}
F^{fk^*l}_{s^*l'k'^*}
\end{align}
(See appendix \ref{grphHam} for a graphical representation of $B_{\v p}^s$).
One can check that the Hamiltonian (\ref{HPi}) is Hermitian if
$F$ satisfies
\begin{equation}
F^{i^*j^*m^*}_{k^*l^*n^*} = (F^{ijm}_{kln})^*
\label{unit}
\end{equation}   
in addition to (\ref{pent}). Our model is only applicable to topological phases
 satisfying this additional constraint. We believe that this is true much more
generally: only topological phases satisfying the unitarity condition
(\ref{unit}) are physically realizable.

The Hamiltonian (\ref{HPi}) has a number of interesting properties, 
provided that $(F^{ijm}_{kln}$, $d_i)$ satisfy the self-consistency conditions 
(\ref{pent}). It turns out that:
\begin{enumerate}
\item{The $B_{\v p}^s$ and $Q_{\v I}$'s all commute with each other.
Thus the Hamiltonian (\ref{HPi}) is exactly soluble.}
\item{Depending on the choice of the coefficients $a_s$, the system can be
in $N+1$ different quantum phases.}
\item{The choice $a_s=\frac{d_s}{\sum_{i=0}^N d_i^2}$ corresponds to a 
topological phase with a smooth continuum limit. 
The ground state wave function for this parameter choice is
topologically invariant, and obeys the local rules (\ref{topinv}-\ref{fusion}).
It is precisely the wave function $\Phi$, defined on a honeycomb
lattice. Furthermore, $Q_{\v I}$, $B_{\v p}$ are projection operators in this 
case. Thus, the ground state satisfies $Q_{\v I} = B_{\v p} = 1$ for all 
$\v I$, $\v p$, while the excited states violate these constraints.}
\end{enumerate}

The Hamiltonian (\ref{HPi}) with the above choice of $a_s$ provides an exactly
soluble realization of the doubled topological phase described by
$F^{ijm}_{kln}$. We can obtain some intuition for the Hamiltonian
(\ref{HPi}) by considering
the case where $F^{ijm}_{kln}$ is the $6j$ symbol of some group $G$. In this
case, it turns out that $Q_{\v I}$ and $B_{\v p}$ are precisely the
electric charge and magnetic flux operators in the standard lattice
gauge theory with group $G$. Thus, (\ref{HPi}) is the usual Hamiltonian of
lattice gauge theory, except with no electric field term. This is nothing
more than the well-known exactly soluble Hamiltonian of lattice gauge theory.
\cite{W7159,K032} In this way, our construction can be viewed as a natural
generalization of lattice gauge theory.

In this paper, we will focus on the smooth topological phase corresponding to
the parameter choice $a_s=\frac{d_s}{\sum_{i=0}^N d_i^2}$ (see appendix
\ref{grphHam}). However, we would like to mention that the other $N$ quantum
phases also have non-trivial topological (or quantum) order. However, in these
phases, the ground state wave function does not have a smooth continuum limit.
Thus, these are new topological phases beyond those described by continuum
theories. 

\subsection{Quasiparticle excitations}

In this section, we find the quasiparticle
excitations of the string-net Hamiltonian (\ref{HPi}), and calculate their
statistics (e.g. the twists $\theta_{\alpha}$ and the $S$ matrix 
$s_{\alpha\beta}$). We will only consider the topological phase with smooth 
continuum limit. That is, we will choose $a_s=\frac{d_s}{\sum_{i=0}^N d_i^2}$ 
in our lattice model.

Recall that the ground state satisfies $Q_{\v I} = B_{\v p} = 1$ for all
vertices $\v I$, and all plaquettes $\v p$. The quasiparticle excitations 
correspond to violations of these constraints for some local collection of 
vertices and plaquettes. We are interested in the topological properties
(e.g. statistics) of these excitations.

\begin{figure}[tb]
\centerline{
\includegraphics[width=3in]{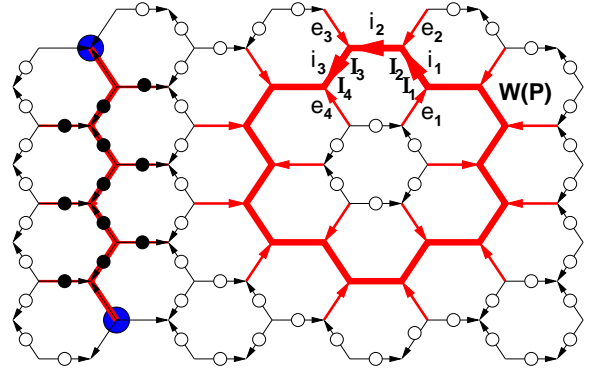}
}
\caption{
Open and closed string operators for the lattice spin model (\ref{HPi}).
Open string operators create quasiparticles at the two ends, as shown on
the left. Closed string operators, as shown on the right, commute with
the Hamiltonian. The closed string operator $W(P)$ only acts 
non-trivially on the spins along the path $P = \v{I_1},\v{I_2}...$ 
(thick line), but its action depends on the spin states on the 
legs (thin lines). The matrix element between an initial 
state $i_1,i_2,...$ and a final state 
$i'_1,i'_2,...$ is $W_{i_1i_2...}^{i'_1i'_2...}(e_1e_2...) =
(F^{e_2i^*_2i_1}_{s^*i'_1i_2'^*} F^{e_3i^*_3i_2}_{s^*i'_2i_3'^*}
...)\cdot (
\frac{\w_{i_1}\w_{s_1}}{\w_{i_1'}}\omega^{i'_1}_{i_1} 
\frac{\w_{i_3}\w_{s_3}}{\w_{i_3'}}\bar{\omega}^{i'_3}_{i_3}
...)$ for a type-$s$ simple string and
$W_{i_1i_2...}^{i'_1i'_2...}(e_1e_2...) = \sum_{\{s_k\}}
(F^{e_2i^*_2i_1}_{s^*_2i'_1i_2'^*} 
F^{e_3i^*_3i_2}_{s^*_3i'_2i_3'^*}
...)\cdot \Tr(
\frac{\w_{i_1}\w_{s_1}}{\w_{i_1'}}
\Omega^{i'_1}_{s_1s_2i_1} 
\del_{s_2s_3}
\text{Id} 
\frac{\w_{i_3}\w_{s_3}}{\w_{i_3'}}
\bar{\Omega}^{i'_3}_{s_3s_4i_3}
...)$ for a general string.
}
\label{quasi}
\end{figure}

We will focus on topologically nontrivial quasiparticles - that is, particles
with nontrivial statistics or mutual statistics. By the analysis in 
\Ref{LWsta}, we know that these types of particles are always created in 
pairs, and that their pair creation operator has a string-like structure, with 
the newly created particles appearing at the ends. (See Fig. \ref{quasi}). 
The position of this string operator is unobservable in the string-net 
condensed state - only the endpoints of the string are observable. Thus the 
two ends of the string behave like independent particles.  

If the two endpoints of the string coincide so that the string forms a loop, 
then the associated closed string operator commutes with the Hamiltonian. This 
follows from the fact that the string is truly unobservable; the action of an 
open string operator on the ground state depends only on its endpoints.

Thus, \emph {each topologically nontrivial quasiparticle is associated with a 
(closed) string operator that commutes with the Hamiltonian}. To find the 
quasiparticles, we need to find these closed string operators.

An important class of string operators are what we will call ``simple'' string 
operators. The defining property of simple string operators is their action on 
the vacuum state. If we apply a type-$s$ simple string operator $W(P)$ to the 
vacuum state, it creates a type-$s$ string along the path of the string, $P$. 
We already have some examples of these operators, namely the magnetic flux 
operators $B^{s}_{\v p}$. When $B^{s}_{\v p}$ acts on the vacuum configuration 
$|0\>$, it creates a type-$s$ string along the boundary of the plaquette 
$\v p$. Thus, we can think of $B^{s}_{\v p}$ as a short type-$s$ simple string 
operator, $W(\partial \v p)$.

We would like to construct simple string operators $W(P)$ for arbitrary 
paths $P = \v{I_1}, ... ,\v{I_N}$ on the honeycomb lattice. Using the 
definition of $B^{s}_{\v p}$ as a guide, we make the following ansatz.
The string operator $W(P)$ only changes the spin states along the path $P$.
The matrix element of a general type-$s$ simple string operator 
$W(P)$ between an initial spin state $i_1,...i_N$ and final spin state 
$i'_1,...i'_N$ is of the form 
\begin{equation}
\label{simpstr}
W_{i_1i_2...i_N}^{i_1'i_2'...i_N'}(e_1e_2...e_N) = 
\left(\prod_{k=1}^{N} F^s_k \right)
\left( \prod_{k=1}^{N} \omega_k \right)
\end{equation}
where $e_1,...,e_N$ are the spin states of the $N$ ``legs'' of $P$ (see Fig. 
\ref{quasi}) and 
\begin{equation}
F^s_k = \left\{
\bmm
F^{e_{k}i_{k}^*i_{k-1}}_{s^*i_{k-1}'i_{k}'^*}, & 
\text{if $P$ turns left at $\v{I_k}$} \\
F^{e_{k}i_{k-1}^*i_{k}}_{si_{k}'i_{k-1}'^*}, & 
\text{if $P$ turns right at $\v{I_k}$}
\emm
\right. 
\end{equation}
\begin{equation}
\label{omijs}
\omega_k = \left\{
\bmm
\frac{\w_{i_k}\w_{s}}{\w_{i_k'}}\omega^{i_{k}'}_{i_k}, &
\text{if $P$ turns right, left at $\v{I_k}$, $\v{I_{k+1}}$} \\
\frac{\w_{i_k}\w_{s}}{\w_{i_k'}}\bar{\omega}^{i_{k}'}_{i_k}, &
\text{if $P$ turns left, right at $\v{I_k}$, $\v{I_{k+1}}$} \\
$1$, & \text{otherwise}
\emm
\right.
\end{equation}
Here, $\om^{i}_{j}$, $\bar{\om}^{i}_{j}$ are two (complex) two index 
objects that characterize the string $W$. 

Note the similarity to the definition of $B^s_{\v p}$. The major difference is 
the additional factor $\prod_{k=1}^{N} \omega_k$. We conjecture that 
$|\omega^{j}_{i}\frac{\w_i\w_s}{\w_j}| = 1$ for a type-$s$ string, 
so $\prod_{k=1}^{N} \omega_k$ is simply a phase factor
that depends on the initial and final spin states 
$i_1$, $i_2$, ..., $i_N$, $i'_1$, $i'_2$, ..., $i'_N$. This phase vanishes 
for paths $P$ that make only left or only right turns, such as plaquette 
boundaries $\partial \v p$. In that case, the definition of $W(P)$ coincides 
with $B^s_{\v p}$.

A straightforward calculation shows that the operator $W(P)$ defined above 
commutes with the Hamiltonian (\ref{HPi}) if 
$\omega^{i}_{j}, \bar{\omega}^{i}_{j}$ satisfy
\begin{eqnarray}
\label{omeq1}
\bar{\omega}^{m}_{j}F^{sl^*i}_{kjm^*}\omega^{l}_{i}
\frac{\w_j \w_s}{\w_m} &=&
\sum_{n=0}^{N}F^{ji^*k}_{s^*nl^*}\omega^{n}_{k}F^{jl^*n}_{ksm^*} 
\nonumber \\
\bar{\omega}^{j}_{i} &=& \sum_{k=0}^N \omega^{k}_{i^*}F^{is^*k}_{i^*sj*}
\end{eqnarray}
The solutions to these equations give all the type-$s$ simple string
operators. 

For example, consider the case of Abelian gauge theory. In this case, the
solutions to (\ref{omeq1}) can be divided into three classes. The first class
is given by $s \neq 0$, $\om^{j}_{i}\frac{\w_i\w_s}{\w_j} =
\bar{\om}^{j}_{i}\frac{\w_i\w_s}{\w_j} = 1$. These string operators create
electric flux lines and the associated quasiparticles are electric charges. In
more traditional nomenclature, these are known as (Wegner-)Wilson loop 
operators \cite{W7159,W7445}. The second class of solutions is given by 
$s = 0$, and 
$\om^{j}_{i}\frac{\w_i\w_s}{\w_j} = (\bar{\om}^{j}_{i}\frac{\w_i\w_s}{\w_j})^*
\neq 1$. These string operators create magnetic flux lines and the associated
quasiparticles are magnetic fluxes. The third class has $s \neq 0$ and
$\om^{j}_{i}\frac{\w_i\w_s}{\w_j} = (\bar{\om}^{j}_{i}\frac{\w_i\w_s}{\w_j})^*
\neq 1$.  These strings create both electric and magnetic flux and the
associated quasiparticles are electric charge/magnetic flux bound states. This
accounts for all the quasiparticles in $(2+1)D$ Abelian gauge theory.
Therefore, all the string operators are simple in this case. 

However, this is not true for non-Abelian gauge theory or other $(2+1)D$ 
topological phases. To compute the quasiparticle spectrum of these more 
general theories, we need to generalize the expression (\ref{simpstr}) for 
$W(P)$ to include string operators that are not simple.

One way to guess the more general expression for $W(P)$ is to consider
products of simple string operators. Clearly, if $W_1(P)$ and $W_2(P)$
commute with the Hamiltonian, then $W(P) = W_1(P) \cdot W_2(P)$ also commutes
with the Hamiltonian. Thus, we can obtain other string operators by taking
products of simple string operators. In general, the resulting operators
are not simple. If $W_1$ and $W_2$ are type-$s_1$ and type-$s_2$ simple
string operators, then the action of the product string on the vacuum state is:
\begin{displaymath}
W(P) |0\> = W_1(P)W_2(P)|0\> = W_1(P) |s_2\> = 
\sum_{s} \del_{ss_1s_2} |s\>
\end{displaymath}
where $|s\>$ denotes the string state with a type-$s$ string along the path 
$P$ and the vacuum everywhere else. If we take products of more then two
simple string operators then the action of the product string on the
vacuum is of the form $W(P)|0\> = \sum_{s} n_{s} |s\>$ where $n_s$ are
some non-negative integers. 

We now generalize the expression for $W(P)$ so that it includes
arbitrary products of simple strings. Let $W$ be a product of simple 
string operators, and let $n_s$ be the non-negative integers characterizing 
the action of $W$ on the vacuum: $W(P)|0\> = \sum_{s} n_{s} |s\>$. 
Then, one can show that the matrix elements of $W(P)$ are always of the 
form
\begin{equation}
\label{strop}
W_{i_1i_2...i_N}^{i_1'i_2'...i_N'}(e_1e_2...e_N) = 
\sum_{\{s_k\}} \left (\prod_{k=1}^{N} F^{s_k}_k \right ) 
\Tr \left(\prod_{k=1}^{N} \Omega^{s_k}_k \right) 
\end{equation}
where
\begin{equation}
\Omega^{s_k}_k = \left\{
\bmm
\frac{\w_{i_k}\w_{s_k}}{\w_{i_k'}} \Omega^{i_{k}'}_{s_{k}s_{k+1}i_k}, & 
\text{if $P$ turns right, left at $\v{I_k}$, $\v{I_{k+1}}$} \\
\frac{\w_{i_k}\w_{s_k}}{\w_{i_k'}}\bar{\Omega}^{i_{k}'}_{s_{k}s_{k+1}i_k}, & 
\text{if $P$ turns left, right at $\v{I_k}$, $\v{I_{k+1}}$} \\ 
\delta_{s_{k}s_{k+1}} \cdot \text{Id}, &
\text{otherwise}
\emm
\right. 
\end{equation}
and $\Omega^{i}_{stj}$,$\bar{\Omega}^{i}_{stj}$ are two $4$ index
objects that characterize the string operator $W$. For any quadruple of string 
types $i,j,s,t$, $(\Omega^{i}_{stj}$, $\bar{\Omega}^{i}_{stj})$ are 
(complex) rectangular matrices of dimension $n_s \times n_t$. Note that 
type-$s_0$ simple string operators correspond to the special case where 
$n_s = \del_{s_0s}$. In this case, the matrices $\Omega^{i}_{stj}, 
\bar{\Omega}^{i}_{stj}$ reduce to complex numbers, and we can identify 
\begin{equation}
\label{Omom}
\Omega^{i}_{stj} = \omega^{i}_{j} \del_{ss_0}\del_{ts_0}
,\ \ \ \ \ \ \ \ \
\bar \Omega^{i}_{stj} = \bar \omega^{i}_{j} \del_{ss_0}\del_{ts_0}
.
\end{equation}
As we mentioned above, products of simple string operators are always of the
form (\ref{strop}). In fact, we believe that \emph{all string operators} are
of this form. Thus, we will use (\ref{strop}) as an ansatz for general string
operators in $(2+1)D$ topological phases. This ansatz is complicated
algebraically, but like the definition of $B^s_{\v p}$, it has a simple
graphical interpretation (see appendix \ref{grphStr}).

A straightforward calculation shows that the closed string $W(P)$ commutes
with the Hamiltonian (\ref{HPi}) if $\Omega$ and $\bar{\Omega}$ satisfy
\begin{eqnarray}
\label{omeq}
\sum_{s=0}^{N}\bar{\Omega}^{m}_{rsj}F^{sl^*i}_{kjm^*}\Omega^{l}_{sti}
\frac{\w_j \w_s}{\w_m} &=&
\sum_{n=0}^{N}F^{ji^*k}_{t^*nl^*}\Omega^{n}_{rtk}F^{jl^*n}_{krm^*} 
\nonumber \\
\bar{\Omega}^{j}_{sti} &=& \sum_{k=0}^N \Omega^{k}_{sti^*}F^{it^*k}_{i^*sj*}
\end{eqnarray}
The solutions $(\Omega_{m},\bar{\Omega}_{m})$ to these equations 
give all the different closed string operators $W_{m}$. However, not all 
of these solutions are really distinct. Notice that two solutions 
$(\Omega_{1}, \bar{\Omega}_{1})$,
$(\Omega_{2}, \bar{\Omega}_{2})$ can be combined to form a new solution
$(\Omega', \bar{\Omega}')$:
\begin{eqnarray}
\Omega'^{j}_{sti} = \Omega^{j}_{1,sti} \oplus \Omega^{j}_{2,sti} 
\nonumber \\
\bar{\Omega}'^{j}_{sti} = \bar{\Omega}^{j}_{1,sti} \oplus 
\bar{\Omega}^{j}_{2,sti} 
\end{eqnarray}
This is not surprising: the string operator $W'$ corresponding to 
$(\Omega', \bar{\Omega}')$ is simply the sum of the two operators
corresponding to $(\Omega_{1,2}, \bar{\Omega}_{1,2})$: $W' = W_{1} + W_{2}$.

Given this additivity property, it is natural to consider the ``irreducible'' 
solutions $(\Omega_{\alpha}, \bar{\Omega}_{\alpha})$ that cannot be written as 
a sum of two other solutions. Only the ``irreducible'' string operators 
$W_{\alpha}$ create quasiparticle-pairs in the usual sense. Reducible string 
operators $W$ create superpositions of different strings - which correspond to 
superpositions of different quasiparticles. \footnote{Note that reducible 
quasiparticles should not be confused with bound states. Indeed, in the case of
Abelian gauge theory, most of the irreducible quasiparticles are bound states 
of electric charges and magnetic fluxes.}

To analyze a topological phase, one only needs to find the irreducible
solutions $(\Omega_{\alpha},\bar{\Omega}_{\alpha})$ to \Eq{omeq}. The number
$M$ of such solutions is always finite. In general, each solution corresponds 
to an irreducible representation of an algebraic object. In the case of lattice
gauge theory, there is one solution for every irreducible representation of the
quantum double $D(G)$ of the gauge group $G$. Similarly, in the case of doubled
Chern-Simons theories there is one solution for each irreducible representation
of a doubled quantum group. 

The structure of these irreducible string operators $W_\alpha$ determines all
the universal features of the topological phase. The number $M$ of
irreducible string operators is the number of different kinds of
quasiparticles. The fusion rules $W_\al W_\bt=\sum_{\ga=1}^M h_{\al\bt}^\ga
W_\ga$ determine how bound states of type-$\al$ and type-$\bt$
quasiparticles can be viewed as a superposition of other types of
quasiparticles. 

The topological properties of the quasiparticles are also easy to
compute. As an example, we now derive two particularly fundamental objects 
that characterize the spins and statistics of quasiparticles: the $M$
twists $\theta_{\alpha}$ and the $M \times M$ S-matrix, $s_{\alpha\beta}$
\cite{V8860,W8951,W8929,Wrig}. 

The twists $\theta_{\alpha}$ are defined to be statistical angles of the
type-$\alpha$ quasiparticles. By the spin-statistics theorem they are closely
connected to the quasiparticle spins $s_\alpha$: $e^{i\theta_{\alpha}} =
e^{2\pi i s_{\alpha}}$. We can calculate $\theta_{\alpha}$ by comparing the 
quantum mechanical amplitude for the following two processes. In the first 
process, we create a pair of quasiparticles $\alpha, \bar{\alpha}$ (from the 
ground state), exchange them, and then annihilate the pair. In the second 
process, we create and then annihilate the pair without any exchange. The ratio
of the amplitudes for these two processes is precisely $e^{i\theta_{\alpha}}$.

The amplitude for each process is given by the expectation value
of the closed string operator $W_{\alpha}$ for a particular path $P$:
\begin{eqnarray}
\mathcal{A}_1 &=& \Big\<\Phi \Big| 
\bmm \includegraphics[height=0.3in]{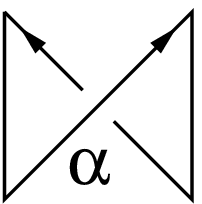} \emm \Big| \Phi \Big\> \\
\mathcal{A}_2 &=& \Big\<\Phi \Big| 
\bmm \includegraphics[height=0.3in]{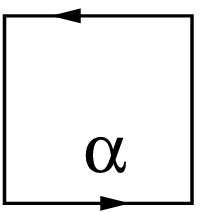} \emm \Big| \Phi \Big\>
\end{eqnarray}
Here, $|\Phi\>$ denotes the ground state of the Hamiltonian (\ref{HPi}).

Let $(\Om_\al,\bar\Om_\al, n_{\al})$ be the irreducible solution 
corresponding to the string operator $W_{\alpha}$. The above two 
amplitudes can be then be expressed in terms of $(\Om_\al,\bar\Om_\al, 
n_{\al})$ (see appendix \ref{grphStr}):
\begin{eqnarray}
\mathcal{A}_1 &=& \sum_{s} d_s^2 \cdot \Tr(\Omega^{0}_{\al,sss^*}) \\
\mathcal{A}_2 &=& \sum_{s} n_{\al,s}d_s
\end{eqnarray}
Combining these results, we find that the twists are given by
\begin{equation}
e^{i\theta_{\alpha}} = \frac{\mathcal{A}_1}{\mathcal{A}_2}
= \frac{\sum_{s} d_s^2 \cdot 
\Tr(\Omega^{0}_{\al,sss^*})}{\sum_{s} n_{\al,s} d_s}
\label{twist}
\end{equation}

Just as the twists $\theta_{\alpha}$ are related to the spin and statistics of 
individual particle types $\alpha$, the elements of the S-matrix, 
$s_{\alpha\beta}$ describe the mutual statistics of two 
particle types $\alpha,\beta$. Consider the following process:
We create two pairs of quasiparticles 
$\alpha, \bar{\alpha}, \beta, \bar{\beta}$, braid $\alpha$ around $\beta$,
and then annihilate the two pairs. The element $s_{\alpha\beta}$
is defined to be the quantum mechanical amplitude $\mathcal{A}$ of this 
process, divided by a proportionality factor $D$ where
$D^2 = \sum_{\alpha} (\sum_{s} n_{\al,s}d_s)^2$. The amplitude
$\mathcal{A}$ can be calculated from the expectation value of 
$W_\alpha, W_\beta$ for two ``linked'' paths $P$:
\begin{equation}
\mathcal{A}=  \Big\<\Phi \Big| 
\bmm \includegraphics[height=0.3in]{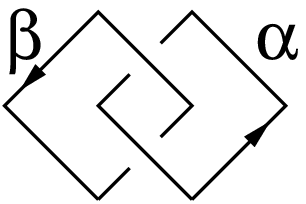} \emm \Big| \Phi \Big\>
\end{equation}
Expressing $\mathcal{A}$ in terms of $(\Om_\al,\bar\Om_\al, 
n_{\al})$, we find
\begin{eqnarray}
s_{\alpha\beta} = \frac{\mathcal{A}}{D} 
= \frac{1}{D}\sum_{ijk}\Tr(\Omega^{k}_{\al,iij^*}) \cdot 
\Tr(\Omega^{k^*}_{\beta,jji^*})d_i d_j
\label{smatrix}
\end{eqnarray}

\section{String-net condensation in $(3+1)$ and higher dimensions}

In this section, we generalize our results to $(3+1)$ and higher dimensions.
We find that there is a one-to-one correspondence between $(3+1)$ (and higher)
dimensional string-net condensates and mathematical objects known as
``symmetric tensor categories.'' \cite{Kas95} The low energy effective theories
for these states are gauge theories coupled to bosonic or fermionic charges. 

Our approach is based on the exactly soluble lattice spin Hamiltonian
\Eq{HPi}. In that model, the spins live on the links of the honeycomb 
lattice. However, the choice of lattice was somewhat arbitrary:
we could equally well have chosen any trivalent lattice in two dimensions. 

\begin{figure}[tb]
\centerline{
\includegraphics[width=3in]{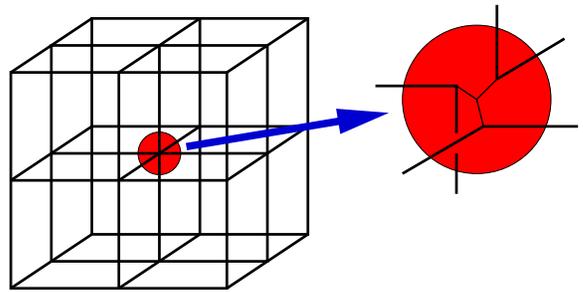}
}
\caption{
A three dimensional trivalent lattice, obtained by splitting
the sites of the cubic lattice. We replace each vertex of the cubic lattice
with $4$ other vertices as shown above.
} 
\label{3Dtri}
\end{figure}

Trivalent lattices can also be constructed in three and higher dimensions.
For example, we can create a space-filling trivalent lattice in three
dimensions, by ``splitting'' the sites of the cubic lattice 
(see Fig. \ref{3Dtri}). Consider the spin Hamiltonian \Eq{HPi} for this 
lattice, where $\v I$ runs over all the vertices of the lattice, and $\v p$ 
runs over all the ``plaquettes'' (that is, the closed loops that correspond to 
plaquettes in the original cubic lattice).

This model is a natural candidate for string-net condensation in three 
dimensions. Unfortunately, it turns out that the Hamiltonian (\ref{HPi}) is not
exactly soluble on this lattice. The magnetic flux operators $B^{s}_{\v p}$ do 
not commute in general. 

\begin{figure}[tb]
\centerline{
\includegraphics[width=3in]{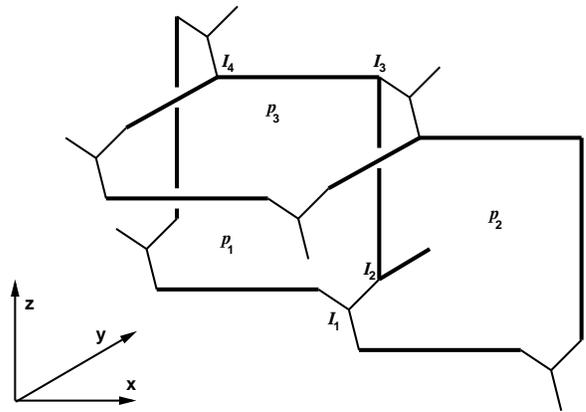}
}
\caption{
Three plaquettes demonstrating the two fundamental differences between higher
dimensional trivalent lattices and the honeycomb lattice. The plaquettes 
$\v{p_1}$, $\v{p_2}$ lie in the $xz$ plane, while $\v{p_3}$ is oriented in the 
$xy$ direction. Notice that $\v{p_1}$ and $\v{p_2}$ share three vertices, 
$\v{I_1}$, $\v{I_2}$, $\v{I_3}$. Also, notice that the plaquette boundaries 
$\partial\v{p_1}$ and $\partial\v{p_3}$ intersect only at the line segment 
$\v{I_3}\v{I_4}$. The boundary $\partial \v{p_1}$ makes a left turn at 
$\v{I_3}$, and a right turn at $\v{I_4}$. Thus, viewed from far away, these 
two plaquette boundaries intersect exactly once (unlike the pair $\partial 
\v{p_1}$ and $\partial \v{p_2}$).
}
\label{3Dplaq}
\end{figure}

This lack of commutativity originates from two differences between the 
plaquettes in the honeycomb lattice and in higher dimensional trivalent 
lattices. The first difference is that in the honeycomb lattice, neighboring 
plaquettes always share precisely two vertices, while in higher dimensions the 
boundary between plaquettes can contain three or more vertices 
(see Fig. \ref{3Dplaq}). The existence of these interior vertices has the 
following consequence. Imagine we choose orientation conventions for each 
vertex, so that we have a notion of ``left turns'' and ``right turns'' for 
oriented paths on our lattice (such an orientation convention can be obtained
by projecting the $3D$ lattice onto a $2D$ plane - as in Fig. 
\ref{3Dplaq}). Then, no matter how we assign these orientations the plaquette 
boundaries will always make both left and right turns. Thus, we cannot regard 
the boundaries of the $3D$ plaquettes as small closed strings the
way we did in two dimensions (since small closed strings always make all left 
turns, or all right turns). But the magnetic flux operators $B^{s}_{\v p}$ only
commute if their boundaries are small closed strings. It is this inconsistency 
between the algebraic definition of $B^{s}_{\v p}$ and the topology of the 
plaquettes that leads to the lack of commutativity.

To resolve this problem, we need to define a Hamiltonian using the general 
simple string operators $W(\partial \v p)$ rather then the small closed 
strings $B^{s}_{\v p}$. Suppose $(\om^{i}_{sj}, \bar \om^{i}_{sj})$,
$s = 0,1,...N$ are type-$s$ solutions of (\ref{omeq1}). After picking 
some ``left turn'', ``right turn'' orientation convention at each vertex, we 
can define the corresponding type-$s$ simple string operators $W_s(P)$ as in 
(\ref{simpstr}). Suppose, in addition, that we choose $(\om^{i}_{sj}, \bar 
\om^{i}_{sj})$ so that the string operators satisfy 
$W_{r} \cdot W_{s} = \sum_{t} \del_{rst} W_{t}$ (this property ensures that 
$W_{s}(\partial \v p)$ are analogous to $B^{s}_{\v p}$). Then, a natural higher
dimensional generalization of the Hamiltonian (\ref{HPi}) is 
\begin{equation} 
 H= - \sum_{\v I} Q_{\v I}-\sum_{\v p} W_{\v p}, \ \ \ \ \  W_{\v p}= 
\sum_{s=0}^N a_s W_s(\partial \v p)
\label{HPi3}
\end{equation}
For a two dimensional
lattice, the conditions (\ref{omeq1}) are sufficient to guarantee that the
Hamiltonian (\ref{HPi3}) is an exactly soluble realization of a doubled
topological phase. (This is because the plaquette boundaries $\partial \v 
p$ are not linked and hence the $W_s(\partial \v p)$ all commute). However, in 
higher dimensions, one additional constraint is necessary. 

This constraint stems from the second, and perhaps more fundamental, 
difference between $2D$ and higher dimensional lattices. In two dimensions,
two closed curves always intersect an even number of times. For higher 
dimensional lattices, this is not the case. Small closed curves, in particular
plaquette boundaries, can (in a sense) intersect exactly once 
(see Fig. \ref{3Dplaq}). Because of this, the objects 
$\omega^{i}_{jk}$ must satisfy the additional 
relation:
\begin{equation}
\omega^{i}_{jk} = \bar{\omega}^{i}_{kj}
\label{omsym}
\end{equation}
One can show that if this additional constraint is satisfied, then (a) the
higher dimensional Hamiltonian (\ref{HPi3}) is exactly soluble, and (b) the
ground state wave function $\Phi$ is defined by local topological rules
analogous to (\ref{topinv}-\ref{fusion}). This means that (\ref{HPi3})
provides an exactly soluble realization of topological phases in $(3+1)$ and
higher dimensions.

Each exactly soluble Hamiltonian is associated with a solution 
$(F^{ijm}_{kln}, \omega^{i}_{jk},\bar{\omega}^{i}_{jk})$ of 
(\ref{pent}), (\ref{omeq1}), (\ref{omsym}). By analogy with the two 
dimensional case, we conjecture that there is a one-to-one correspondence 
between topological string-net condensed phases in $(3+1)$ or higher 
dimensions, and these solutions. The solutions $(F^{ijm}_{kln}, 
\omega^{i}_{jk}, \bar{\omega}^{i}_{jk})$ correspond to a special class of 
tensor categories - symmetric tensor categories. \cite{Kas95} Thus, just as
tensor categories are the mathematical objects underlying string condensation 
in $(2+1)$ dimensions, symmetric tensor categories are 
fundamental to string condensation in higher dimensions.
 
There are relatively few solutions to (\ref{pent}), (\ref{omeq1}), 
(\ref{omsym}). 
Physically, this is a consequence of the restrictions on quasiparticle 
statistics in $3$ or higher dimensions. Unlike in two dimensions, higher 
dimensional quasiparticles necessarily have trivial mutual statistics, and must
be either bosonic or fermionic. From a more mathematical point of view, the 
scarcity of solutions is a result of the symmetry condition (\ref{omsym}). 
Doubled topological phases, such as Chern-Simons theories, typically fail to 
satisfy this condition. 

However, gauge theories do satisfy the symmetry condition (\ref{omsym}) and
therefore do correspond to higher dimensional string-net condensates.  Recall
that the gauge theory solution to (\ref{pent}) is obtained by  (a) letting the
string-type index $i$ run over the irreducible representations of the gauge
group, (b) letting the numbers $d_i$ be the dimensions of the representations,
and (c) letting the $6$ index object $F^{ijm}_{kln}$ be the $6j$ symbol of the
group. One can check that this also provides a solution to (\ref{omeq1}),
(\ref{omsym}), if we set $\om^{i}_{jk}\frac{\w_j\w_k}{\w_i} = -1$ when $j = k$
and the invariant tensor in $k \otimes k \otimes i^*$ is antisymmetric in 
the first two indices, and $\om^{i}_{jk}\frac{\w_j\w_k}{\w_i} = 1$ 
otherwise.  This result is to be expected, since the string-net picture 
of gauge theory (section II) is valid in any number of dimensions. Thus, 
it is not surprising that gauge theories can emerge from higher 
dimensional string-net condensation.

There is another class of higher dimensional string-net condensed phases
that is more interesting. The low energy effective theories for these
phases are variants of gauge theories. Mathematically, they are obtained by
twisting the usual gauge theory solution by
\begin{equation}
\tilde{\om}^{i}_{jk} = \om^{i}_{jk} \cdot (-1)^{P(j)P(k)}
\end{equation}
Here $P(i)$ is some assignment of parity (``even'', or ``odd'') to 
each representation $i$. The assignment must be self-consistent in the sense 
that the tensor product of two representations with the same (different)
parity, decomposes into purely even (odd) representations. If all the 
representations are assigned an even parity - that is $P(i)$ is trivial - then 
the twisted gauge theory reduces to standard gauge theory.

The major physical distinction between twisted gauge theories and
standard gauge theories is the quasiparticle spectrum. In standard gauge
theory, the fundamental quasiparticles are the electric charges created by
the $N+1$ string operators $W_i$. These quasiparticles are all bosonic. In
contrast, in twisted gauge theories, all the quasiparticles corresponding to
``odd'' representations $i$ are fermionic. 

In this way, higher dimensional string-net condensation naturally gives rise
to \emph{both} emerging gauge bosons and emerging fermions. This feature 
suggests that gauge interactions and Fermi statistics may be intimately 
connected. The string-net picture may be the bridge between these two 
seemingly unrelated phenomena. \cite{LWsta}

In fact, it appears that gauge theories coupled to fermionic or bosonic charged
particles are the \emph{only} possibilities for higher dimensional string-net 
condensates: mathematical work on symmetric tensor categories suggests
that the only solutions to (\ref{pent}), (\ref{omeq1}), (\ref{omsym}) are
those corresponding to gauge theories and twisted gauge theories. \cite{EG0337}

We would like to point out that $(3+1)$ dimensional string-net condensed
states also exhibit membrane condensation. These membrane operators are
entirely analogous to the string operators. Just as open string operators
create charges at their two ends, open membrane operators create magnetic 
flux 
loops along their boundaries. Furthermore, just as string condensation makes
the string unobservable, membrane condensation leads to the unobservability
of the membrane. Only the boundary of the membrane - the magnetic flux loop - 
is observable.
  
\section{Examples}

\subsection{$N=1$ string model}
\label{Z2SemI}

We begin with the simplest string-net model. In the notation from section III, 
this model is given by 
\begin{enumerate}
\item{Number of string types: $N=1$}
\item{Branching rules: $\emptyset$ (no branching)} 
\item{String orientations: $1^*=1$. }
\end{enumerate}
In other words, the string-nets in this model contain one unoriented string 
type and have no branching. Thus, they are simply closed loops. (See 
Fig. \ref{strgauge}a).

We would like to find the different topological phases that can emerge from 
these closed loops. According to the discussion in section IV, each phase is 
captured by a fixed-point wave function, and each fixed-point wave function is 
specified by local rules (\ref{topinv}-\ref{fusion}) that satisfy the 
self-consistency relations (\ref{pent}). It 
turns out that (\ref{pent}) have only two solutions in this case (up to 
rescaling):

\begin{eqnarray}
d_{0} &=& 1 \nonumber\\
d_{1} &=& F^{110}_{110} = \pm 1 \nonumber \\
F^{000}_{000}&=& 
F^{101}_{101}=
F^{011}_{011}=1 \nonumber \\
F^{000}_{111}&=&
F^{110}_{001}=
F^{101}_{010}=
F^{011}_{100}=1
\label{Z2semsol}
\end{eqnarray}
where the other elements of $F$ all vanish. The corresponding local
rules (\ref{topinv}-\ref{fusion}) are:
\begin{align}
 \Phi
\bpm \includegraphics[height=0.23in]{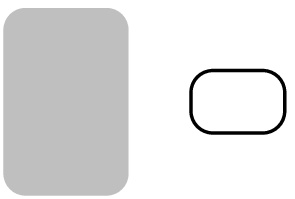} \epm  =&
\pm \Phi 
\bpm \includegraphics[height=0.23in]{X0.eps} \epm \nonumber \\
 \Phi
\bpm \includegraphics[height=0.23in]{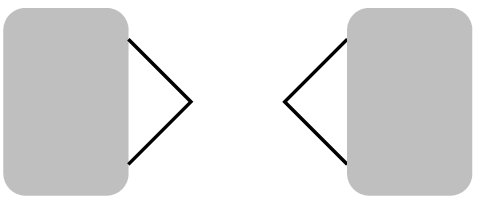} \epm  =&
\pm \Phi 
\bpm \includegraphics[height=0.23in]{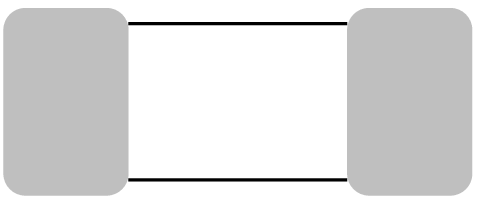} \epm
\end{align}
We have omitted those rules that can be derived from topological invariance
(\ref{topinv}).

The fixed-point wave functions $\Phi_{\pm}$ satisfying these rules are given by
\begin{equation}
\Phi_{\pm}(X) = (\pm 1)^{X_c}
\end{equation}
where $X_c$ is the number of disconnected components in the string 
configuration $X$.

The two fixed-point wave functions $\Phi_{\pm}$ correspond to two simple 
topological phases. As we will see, $\Phi_{+}$ corresponds to $Z_2$ gauge 
theory, while $\Phi_{-}$ is a $U(1) \times U(1)$ Chern-Simons theory.
(Actually, other topological phases can emerge from closed loops - such as in 
\Ref{F0160,FNS0311,FNS0320}. However, we regard these phases as emerging from 
more complicated string-nets. The closed loops organize into these effective 
string-nets in the infrared limit).

\begin{figure}[tb]
\centerline{
\includegraphics[width=2.5in]{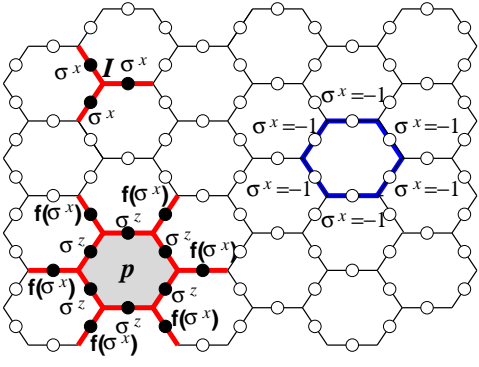}
}
\caption{
The Hamiltonians (\ref{Z2}), (\ref{sem}), realizing the two $N=1$ 
string-condensed phases. Each circle denotes a spin-1/2 spin. The links with 
$\si^{x} = -1$ are thought of as being occupied by a type-$1$ string, while the
links with $\si^{x} = +1$ are regarded as empty. The electric charge term acts 
on the three legs of the vertex $\v I$ with $\si^{x}$. The magnetic energy term
acts on the $6$ edges of the plaquette $\v p$ with $\si^{z}$, and acts on the 
$6$ legs of $\v p$ with an operator of the form $f(\si^{x})$. For the $Z_2$ 
phase, $f = 1$, while for the Chern-Simons phase, $f(x) = i^{(1-x)/2}$.
}
\label{Z2Ham}
\end{figure}

The exactly soluble models (\ref{HPi}) realizing these two phases can be
written as spin $1/2$ systems with one spin on each link of the 
honeycomb lattice (see Fig. \ref{Z2Ham}). We regard a link with 
$\sigma^{x} = -1$ as being occupied by a type-$1$ string, and the state 
$\sigma^{x} = +1$ as being unoccupied (or equivalently, occupied by a type-$0$ 
or null string). The Hamiltonians for the two phases are of the form
\begin{displaymath} 
H_{\pm}= - \sum_{\v I} Q_{\v I,\pm}-\sum_{\v p} B_{\v p,\pm}
\end{displaymath}
The electric charge term is the same for both phases (since it only depends
on the branching rules):
\begin{equation}
Q_{\v I,\pm} = \frac{1}{2}(1 +\prod_{\text{legs of } \v I} \sigma_{\v i}^{x})
\end{equation}
The magnetic terms for the two phases are
\begin{eqnarray}
B_{\v p,\pm} &=& \frac{1}{2}(B_{\v p,\pm}^{0} \pm B_{\v p,\pm}^{1})  \\
&=&\frac{1}{2}\left (1 \pm \prod_{\text{edges of }\v p} \si^z_{\v j} \cdot
\prod_{\text{legs of }\v p} (\sqrt{\pm 1})^{\frac{1-\si^{x}_{\v j}}{2}}
\right) P_{\v p}
\nonumber 
\end{eqnarray}
where $P_{\v p}$ is the projection operator 
$P_{\v p} = \prod_{\v I \in \v p} Q_{\v I}$. The projection operator 
$P_{\v p}$ can be omitted without affecting the physics (or the exact 
solubility of the Hamiltonian). We have included it only to be consistent with 
(\ref{HPi}). If we omit this term, the Hamiltonian for the first phase
($\Phi_{+}$) reduces to the usual exactly soluble Hamiltonian of $Z_2$ lattice
gauge theory (neglecting numerical factors):
\begin{equation}
H_{+} = - \sum_{\v I} \prod_{\text{legs of } \v I} \si_{\v i}^{x}
-\sum_{\v p}\prod_{\text{edges of }\v p} \si^z_{\v j}
\label{Z2}
\end{equation}
The Hamiltonian for the second phase,
\begin{equation}
H_{-} = - \sum_{\v I} \prod_{\text{legs of } \v I} \si_{\v i}^{x}
+\sum_{\v p}(\prod_{\text{edges of }\v p} \si^z_{\v j})
(\prod_{\text{legs of }\v p} i^{\frac{1-\si^{x}_{\v j}}{2}})
\label{sem}
\end{equation} 
is less familiar. However, one can check that in both cases, the Hamiltonians 
are exactly soluble and the two ground state wave functions are precisely 
$\Phi_{\pm}$ (in the $\si^{x}$ basis). 

Next we find the quasiparticle excitations for the two phases, and
the corresponding S-matrix and twists $\th_\al$.

In both cases, equation (\ref{omeq}) has 4 irreducible
solutions $(n_{\al,s}, \Om^{ij}_{\al,st}, \bar \Om^{ij}_{\al,st})$, 
$\al=1,2,3,4$ - corresponding to 4 quasiparticle types. For the first phase 
($\Phi_{+}$) these solutions are given by:
\begin{displaymath}
\begin{array}{llll}
n_{1,0} = 1, & n_{1,1} = 0, & \Omega^{0}_{1,000} =1,& \Omega^{1}_{1,001} =1 \\ 
n_{2,0} = 0, & n_{2,1} = 1, & \Omega^{1}_{2,110} =1,& \Omega^{0}_{2,111} =1 \\
n_{3,0} = 1, & n_{3,1} = 0, & \Omega^{0}_{3,000} =1,& \Omega^{1}_{3,001} =-1 \\
n_{4,0} = 0, & n_{4,1} = 1, & \Omega^{1}_{4,110} =1,& \Omega^{0}_{4,111} =-1 
\end{array}
\end{displaymath}
The other elements of $\Omega$ vanish. In all cases $\bar{\Omega}$ =
$\Omega$.

\begin{figure}[tb]
\centerline{
\includegraphics[width=2.5in]{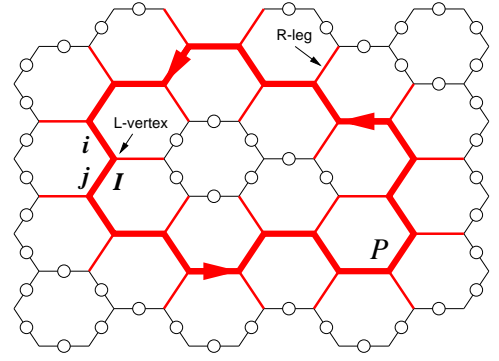}
}
\caption{
A closed string operator $W(P)$ for the two models (\ref{Z2}),(\ref{sem}).
The path $P$ is drawn with a thick line, while the legs are drawn with
thin lines. The action of the string operators (\ref{Z2str}),(\ref{semstr})
on the legs is different for legs that branch to the right of $P$, ``R-legs'', 
and legs that branch to the left of $P$, ``L-legs.'' Similarly, we distinguish 
between ``R-vertices'' and ``L-vertices'' which are ends of ``R-leg'' and
``L-leg'' respectively.
}
\label{Z2quasi}
\end{figure}

The corresponding string operators for a path $P$ are
\begin{eqnarray}
W_{1} &=& \text{Id} \nonumber \\
W_{2} &=& \prod_{\text{edges of } P} \si^z_{\v j} \nonumber \\
W_{3} &=& \prod_{\text{R-legs}} \si^x_{\v k} \nonumber \\
W_{4} &=& \prod_{\text{edges of } P} \si^z_{\v j} \prod_{\text{R-legs}} 
\si^x_{\v k} 
\label{Z2str}
\end{eqnarray}
where the ``R-legs'' $\v k$ are the legs that are to the right of $P$.
(See Fig. \ref{Z2quasi}). Technically, we should multiply these string 
operators by an additional projection operator $\prod_{\v I \in P} Q_{\v I}$, 
in order to be consistent with the general result (\ref{strop}). However, we 
will neglect this factor since it doesn't affect the physics.

Once we have the string operators, we can easily calculate the 
twists and the S-matrix. We find:
\begin{eqnarray}
\begin{array}{llll}
e^{i\th_{1}} = 1, & e^{i\th_{2}} = 1, & e^{i\th_{3}}=1,& e^{i\th_{4}}=-1
\end{array} \\
S = \frac{1}{2}\left ( \begin{array}{cccc}
    1 & 1 & 1 & 1 \\
    1 & 1 & -1 &-1 \\
    1 & -1 & 1 & -1\\
    1 & -1 & -1 & 1 \\
        \end{array} \right )
\end{eqnarray}
This is in agreement with the twists and S-matrix for $Z_2$ gauge theory:
$W_1$ creates trivial quasiparticles, $W_2$ creates magnetic fluxes, 
$W_3$ creates electric charges, $W_4$ creates electric/magnetic bound states.

In the second phase ($\Phi_{-}$), we find
\begin{displaymath}
\begin{array}{llll}
n_{1,0} = 1, & n_{1,1} = 0, & \Omega^{0}_{1,000} = 1, &\Omega^{1}_{1,001} =1 
\\ 
n_{2,0} = 0, & n_{2,1} = 1, & \Omega^{1}_{2,110} = 1, &\Omega^{0}_{2,111} = i 
\\
n_{3,0} = 0, & n_{3,1} = 1, & \Omega^{1}_{3,110} = 1, &\Omega^{0}_{3,111} =-i 
\\
n_{4,0} = 1, & n_{4,1} = 0, & \Omega^{0}_{4,000} = 1, &\Omega^{1}_{4,001} =-1 
\\

\end{array}
\end{displaymath}
Once again, the other elements of $\Omega$ vanish. Also, in all cases,
$\bar{\Omega} = \Omega^*$. The corresponding string operators for a path
$P$ are
\begin{eqnarray}
W_{1} &=& \text{Id} \nonumber \\
W_{2} &=& \prod_{\text{edges of } P} \si^z_{\v j} \prod_{\text{R-legs}} 
 i^{\frac{1-\si^{x}_{\v j}}{2}} \prod_{\text{L-vertices}} (-1)^{s_{\v I}} 
\nonumber \\
W_{3} &=& \prod_{\text{edges of } P} \si^z_{\v j} \prod_{\text{R-legs}} 
(-i)^{\frac{1-\si^{x}_{\v j}}{2}}\prod_{\text{L-vertices}} 
(-1)^{s_{\v I}} \nonumber \\
W_{4} &=& \prod_{\text{R-legs}} \si^x_{\v j} 
\label{semstr}
\end{eqnarray}
where the ``L-vertices'' $\v I$ are the vertices of $P$ adjacent to legs that 
are to the left of $P$. The exponent $s_{\v I}$ is defined by 
$s_{\v I} = \frac{1}{4}(1-\si^{x}_{\v i})(1+\si^{x}_{\v j})$, where
$\v i$, $\v j$ are the links just before and just after the vertex $\v I$,
along the path $P$. (See Fig. \ref{Z2quasi}).

We find the twists and S-matrix are
\begin{eqnarray}
\begin{array}{llll}
e^{i\th_{1}} = 1, & e^{i\th_{2}} = i, & e^{i\th_{3}}=-i,& e^{i\th_{4}}=1
\end{array} \\
S = \frac{1}{2}\left ( \begin{array}{cccc}
    1 & 1 & 1 & 1 \\
    1 & -1 & 1 &-1 \\
    1 & 1 & -1 & -1\\
    1 & -1 & -1 & 1 \\
        \end{array} \right ) 
\end{eqnarray}
We see that $W_1$ creates trivial quasiparticles, $W_2$, $W_3$ create semions 
with opposite chiralities and trivial mutual statistics, and $W_4$ creates
bosonic bound states of the semions.  These results agree with the  
$U(1) \times U(1)$ Chern-Simons theory 
\begin{equation}
L=\frac{1}{4\pi} K_{IJ} a_{I\mu}\prt_\nu a_{J\la}\eps^{\mu\nu\la},\ \ \ \
\ \ \
I,J=1,2 
\end{equation}
with $K$-matrix
\begin{equation}
 K=\bpm 2 & 0 \\ 0 & -2 \epm
\end{equation}
Thus the above $U(1)\times U(1)$ Chern-Simons theory is the low energy
effective theory of the second exactly soluble model (with $d_1=-1$).

Note that the $Z_2$ gauge theory from the first exactly soluble model 
(with $d_1=1$) can also be viewed as a $U(1) \times U(1)$ Chern-Simons theory
with $K$-matrix \cite{HOS0427}
\begin{equation}
 K=\bpm 0 & 2 \\ 2 & 0 \epm
\end{equation}

\subsection{$N=1$ string-net model}
\label{YLmodelI}
The next simplest string-net model also contains only one oriented string type 
- but with branching. Simple as it is, we will see that this model contains 
non-Abelian anyons and is theoretically capable of universal fault tolerant 
quantum computation \cite{FLZ0205}. Formally, the model is defined by
\begin{enumerate}
\item{Number of string types: $N=1$}
\item{Branching rules: $\{\{1,1,1\}\}$} 
\item{String orientations: $1^*=1$}. 
\end{enumerate}
The string-nets are unoriented trivalent graphs. To find the topological
phases that can emerge from these objects, we solve the self-consistency
relations (\ref{pent}). We find two sets of self-consistent rules:
\begin{align}
\label{YLsol}
 \Phi
\bpm \includegraphics[height=0.23in]{Xl.eps} \epm  =& 
\ga_\pm  \cdot \Phi 
\bpm \includegraphics[height=0.23in]{X0.eps} \epm \nonumber \\
 \Phi
\bpm \includegraphics[height=0.23in]{XijklX.eps} \epm =& 
\ga_\pm^{-1} \cdot \Phi 
\bpm \includegraphics[height=0.23in]{XijX.eps} \epm  \nonumber \\ 
+& \ga_\pm^{-1/2} \cdot \Phi 
\bpm \includegraphics[height=0.23in]{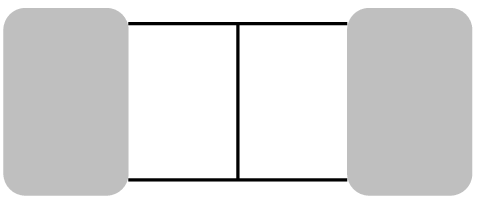} \epm \nonumber \\
 \Phi
\bpm \includegraphics[height=0.23in]{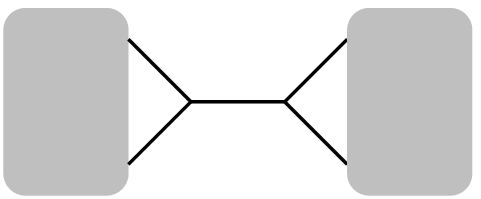} \epm  =& 
\ga_\pm^{-1/2} \cdot \Phi 
\bpm \includegraphics[height=0.23in]{XijklX.eps} \epm \nonumber \\
-& \ga_\pm^{-1} \cdot \Phi 
\bpm \includegraphics[height=0.23in]{XijklnUX.eps} \epm
\end{align}
where $\ga_{\pm} = \frac{1 \pm \sqrt{5}}{2}$.
(Once again, we have omitted those rules that can be derived from topological
invariance). Unlike the previous case, there is no closed form expression for 
the wave function amplitude.

Note that the second solution, $d_{1} = \frac{1 - \sqrt 5}{2}$ does not
satisfy the unitarity condition (\ref{unit}). Thus, only the first solution 
corresponds to a physical topological phase. As we will see, this phase is 
described by an $SO_{3}(3) \times SO_{3}(3)$ Chern-Simons theory.

As before, the exactly soluble realization of this phase (\ref{HPi}) is a 
spin-1/2 model with spins on the links of the honeycomb lattice. We regard
a link with $\si^x = -1$ as being occupied by a type-$1$ string, and a
link with $\si^x = 1$ as being unoccupied (or equivalently occupied by a
type-$0$ string). However, in this case we will not explicitly rewrite 
(\ref{HPi}) in terms of Pauli matrices, since the resulting expression is quite
complicated.

We now find the quasiparticles. These correspond to irreducible solutions of 
(\ref{omeq}). For this model, there are $4$ such solutions, corresponding to 
$4$ quasiparticles:
\begin{align}
1:\ \ & n_{1,0} = 1, \ n_{1,1} = 0, \ \Omega^{0}_{1,000} = 1, \ 
\Omega^{1}_{1,001} =1 \nonumber \\ 
2:\ \ & n_{2,0} = 0, \ n_{2,1} = 1, \ \Omega^{1}_{2,110} = 1, 
\\
& \ \ \ \Omega^{0}_{2,111} = -\gamma_+^{-1} e^{\pi i/5}, \ 
\Omega^{1}_{2,111} =
\gamma_+^{-1/2}e^{3\pi i/5} 
\nonumber \\
3:\ \ & n_{3,0} = 0, \ n_{3,1} = 1, \ \Omega^{1}_{3,110} = 1, 
\nonumber \\
& \ \ \ \Omega^{0}_{3,111} = -\gamma_+^{-1} e^{-\pi i/5}, \ 
\Omega^{1}_{3,111} = \gamma_+^{-1/2}e^{-3\pi i/5} 
\nonumber \\
4:\ \ & n_{4,0} = 1, \ n_{4,1} = 1, \ \Omega^{0}_{4,000} = 1, \ 
\Omega^{1}_{4,110} = 1, \nonumber \\
& \ \ \ \Omega^{1}_{4,001} = -\gamma_+^{-2}, \ \Omega^{0}_{4,111} = 
\gamma_+^{-1}, \ \Omega^{1}_{4,111} = \gamma_+^{-5/2}, 
\nonumber\\
& \ \ \ \Omega^{1}_{4,101} = (\Omega^{1}_{4,011})^* = 
\gamma_+^{-11/4} (2 - e^{3\pi i/5} +\gamma_+ e^{-3\pi i/5}).
\nonumber 
\end{align}
In all cases, $\bar{\Omega} = \Omega^*$.

We can calculate the twists and the S-matrix. We find:
\begin{align}
&\begin{array}{llll}
e^{i\th_{1}} = 1, & e^{i\th_{2}} = e^{-4\pi i/5}, & 
e^{i\th_{3}}= e^{4\pi i/5}, & e^{i\th_{4}}=1
\end{array} \\
&S = \frac{1}{1+\ga^2}\left ( \begin{array}{cccc}
    1 & \ga & \ga & \ga^2 \\
    \ga & -1 & \ga^2 &-\ga \\
    \ga & \ga^2 & -1 & -\ga \\
    \ga^2 & -\ga & -\ga & 1 \\
        \end{array} \right ) 
\end{align}
We conclude that $W_1$ creates trivial quasiparticles, $W_2$, $W_3$
create (non-Abelian) anyons with opposite chiralities, and $W_4$ creates 
bosonic bound states of the anyons. These results agree with
$SO_{3}(3) \times SO_{3}(3)$ Chern-Simons theory, the so-called 
doubled ``Yang-Lee'' theory. 

Researchers in the field of quantum computing have shown that the Yang-Lee 
theory can function as a universal quantum computer - via manipulation of 
non-Abelian anyons. \cite{FLZ0205} Therefore, the spin-1/2 Hamiltonian 
(\ref{HPi}) associated with (\ref{YLsol}) is a theoretical realization
of a universal quantum computer. While this Hamiltonian may be too
complicated to be realized experimentally, the string-net picture suggests
that this problem can be overcome. Indeed, the string-net picture 
suggests that \emph{generic} spin Hamiltonians with a trivalent graph 
structure will exhibit a Yang-Lee phase. Thus, much simpler spin-1/2 
Hamiltonians may be capable of universal fault tolerant quantum 
computation.

\subsection{$N = 2$ string-net models}
In this section, we discuss two $N=2$ string-net models. The first model
contains one oriented string and its dual. In the notation from section III, it
is given by
\begin{enumerate}
\item{Number of string types: $N=2$}
\item{Branching rules: $\{\{1,1,1\}, \{2,2,2\}\}$} 
\item{String orientations: $1^*=2$, $2^* = 1$}. 
\end{enumerate}
The string-nets are therefore oriented trivalent graphs with $Z_3$ branching
rules. The string-net condensed phases correspond to solutions of 
(\ref{pent}). Solving these equations, we find two sets of self-consistent
local rules:
\begin{align}
 \Phi
\bpm \includegraphics[height=0.23in]{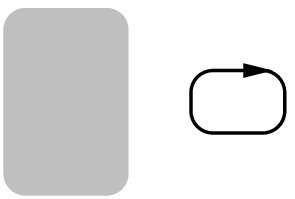} \epm  =& 
\pm \Phi 
\bpm \includegraphics[height=0.23in]{X0.eps} \epm \nonumber \\
\Phi
\bpm \includegraphics[height=0.23in]{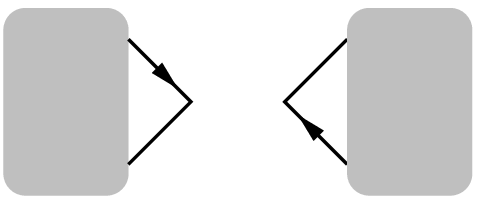} \epm =& 
\pm \Phi 
\bpm \includegraphics[height=0.23in]{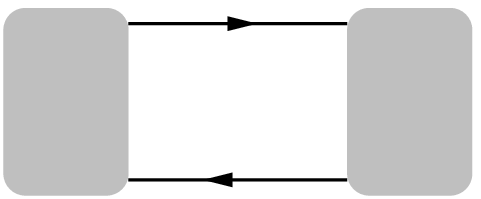} \epm  \nonumber \\
\Phi
\bpm \includegraphics[height=0.23in]{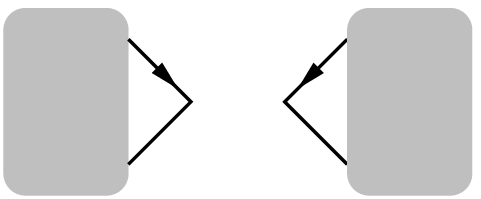} \epm  =& 
\sqrt{\pm 1} \cdot \Phi 
\bpm \includegraphics[height=0.23in]{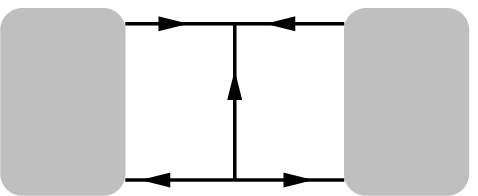} \epm 
\end{align}
The corresponding fixed-point wave functions $\Phi_{\pm}$ are given by
\begin{equation}
\Phi_{\pm}(X) = (\sqrt{\pm 1})^{2X_{c}-X_{v}/2}
\end{equation}
where $X_c$, $X_v$, are the number of connected components, and vertices,
respectively in the string-net configuration $X$. As before, we can construct
an exactly soluble Hamiltonians, find the quasiparticles for the two theories 
and compute the twists and S-matrices. We find that the first theory 
$\Phi_{+}$ is described by a $Z_3$ gauge theory, while the second theory 
$\Phi_{-}$ is described by a $U(1) \times U(1)$ Chern-Simons theory with 
$K$-matrix
\begin{equation*}
 K=\bpm 3 & 0 \\ 0 & -3 \epm
\end{equation*}
Both theories have $3^2 = 9$ elementary quasiparticles. In the case of $Z_3$,
these quasiparticles are electric charge/magnetic flux bound states 
formed 
from the $3$ types of electric charges and $3$ types of magnetic fluxes. In
the case of the Chern-Simons theory, the quasiparticles are
bound states of the two fundamental anyons with statistical angles $\pm \pi/3$.

The final example we will discuss contains two unoriented strings. Formally it
is given by
\begin{enumerate}
\item{Number of string types: $N=2$}
\item{Branching rules: $\{\{1,2,2\}, \{2,2,2\}\}$} 
\item{String orientations: $1^*=1$, $2^* = 2$}. 
\end{enumerate}
The string-nets are unoriented trivalent graphs, with edges labeled with
$1$ or $2$. We find that there is only one set of self-consistent local
rules:
\begin{align}
 \Phi
\bpm \includegraphics[height=0.23in]{Xil0.eps} \epm  =&
d_i \cdot \Phi 
\bpm \includegraphics[height=0.23in]{X0.eps} \epm \nonumber \\
 \Phi
\bpm \includegraphics[height=0.23in]{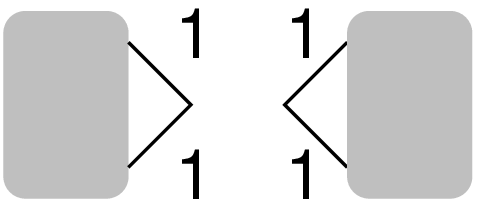} \epm  =&
\Phi 
\bpm \includegraphics[height=0.23in]{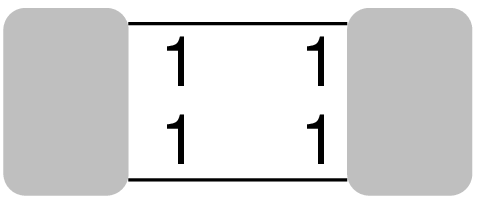} \epm \nonumber \\
 \Phi
\bpm \includegraphics[height=0.23in]{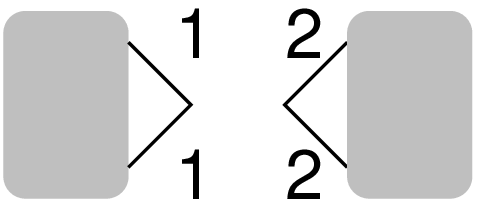} \epm  =&
\Phi 
\bpm \includegraphics[height=0.23in]{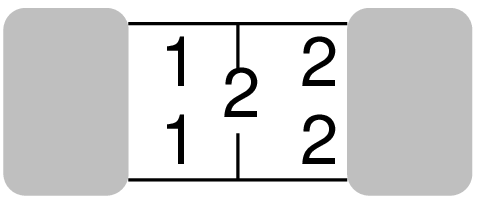} \epm \nonumber \\
 \Phi
\bpm \includegraphics[height=0.23in]{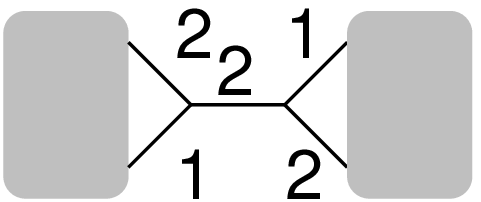} \epm  =&
 \Phi 
\bpm \includegraphics[height=0.23in]{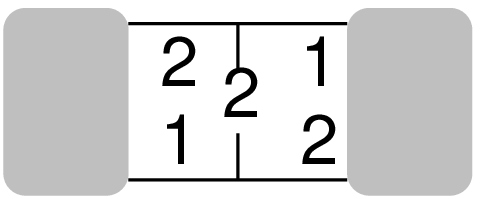} \epm \nonumber \\
 \Phi
\bpm \includegraphics[height=0.23in]{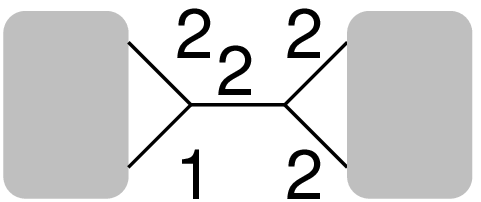} \epm  =&
-\Phi 
\bpm \includegraphics[height=0.23in]{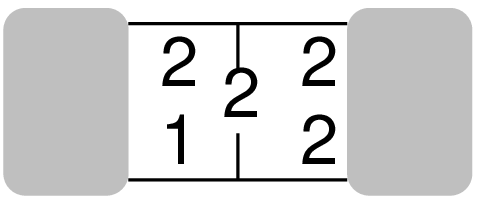} \epm \nonumber \\
 \Phi
\bpm \includegraphics[height=0.23in]{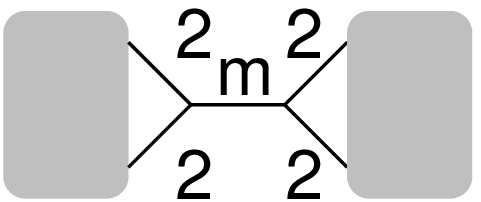} \epm  =&
\sum_{n=0}^2F^{22m}_{22n} \cdot \Phi 
\bpm \includegraphics[height=0.23in]{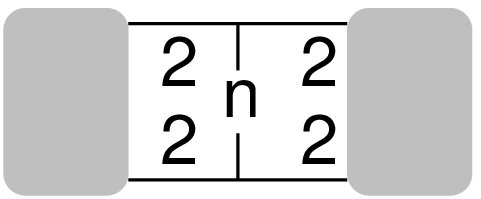} \epm
\end{align}
where $d_0 = d_1 = 1$, $d_2 = 2$, and $F^{22m}_{22n}$ is the matrix
\begin{equation*}
 F^{22m}_{22n} =\bpm \frac{1}{2} & \frac{1}{2} & \frac{1}{\sqrt{2}} \\
                   \frac{1}{2}     & \frac{1}{2} & -\frac{1}{\sqrt{2}} \\
		 \frac{1}{\sqrt{2}} & -\frac{1}{\sqrt{2}} & 0 \epm
\end{equation*}

If we construct the Hamiltonian (\ref{HPi}), we find that it
is equivalent to the standard exactly lattice gauge theory Hamiltonian 
\cite{K032} with gauge group $S_3$ - the permutation group on $3$ objects. One 
can show that this theory contains $8$ elementary quasiparticles 
(corresponding to the  $8$ irreducible representations of the quantum double 
$D(S_3)$). These quasiparticles are combinations of the $3$ electric charges 
and $3$ magnetic fluxes. 

\section{Conclusion}
In this paper, we have shown that quantum systems of extended objects
naturally give rise to topological phases. These phases occur when the extended
objects (e.g. string-nets) become highly fluctuating and condense. This
physical picture provides a natural mechanism for the emergence of 
parity invariant topological phases. Microscopic degrees of freedoms (such as 
spins or dimers) can organize into effective extended objects which can then 
condense. We hope that this physical picture may help direct the search for 
topological phases in real condensed matter systems. It would be interesting
to develop an analogous picture for chiral topological phases.

We have also found the fundamental mathematical framework for topological 
phases. We have shown that each $(2+1)$ dimensional doubled topological phase 
is associated with a $6$ index object $F^{ijm}_{kln}$ and a set of real numbers
$d_i$ satisfying the algebraic relations (\ref{pent}). All the universal 
properties of the topological phase are contained in these mathematical objects
(known as tensor categories). In particular, the tensor category directly 
determines the quasiparticle statistics of the associated topological phase
(\ref{twist}, \ref{smatrix}). This mathematical framework may also have 
applications to phase transitions and critical phenomena. Tensor 
categories may characterize transitions between topological phases 
just as symmetry groups characterize transitions between ordered phases.

We have constructed exactly soluble $(2+1)D$ lattice spin Hamiltonians
(\ref{HPi}) realizing each of these doubled topological phases. These models
unify $(2+1)D$ lattice gauge theory and doubled Chern-Simons theory.  
One particular Hamiltonian - a realization of the doubled Yang-Lee theory 
- is a spin 1/2 model capable of fault tolerant quantum computation.

In higher dimensions, string-nets can also give rise to topological phases.
However, the physical and mathematical structure of these phases is more
restricted. On a mathematical level, each higher dimensional string-net
condensate is associated with a special kind of tensor category - a symmetric
tensor category (\ref{omeq1}), (\ref{omsym}). More physically, we have shown
that higher dimensional string-net condensation naturally gives rise to
\emph{both} gauge interactions and Fermi statistics.  Viewed from this
perspective, string-net condensation provides a mechanism for unifying 
gauge interactions and Fermi statistics. It may have applications to high 
energy physics \cite{Wqoem}.

From a more general point of view, all of the phases described by Landau's
symmetry breaking theory can be understood in terms of particle 
condensation.  These phases are classified using group theory and lead to 
emergent gapless scalar bosons \cite{N6080,G6154}, such as phonons, spin waves,
\etc. In this paper, we have shown that there is a much richer class of phase -
arising from the condensation of extended objects. These phases are classified 
using tensor category theory and lead to emergence of Fermi statistics and 
gauge excitations. Clearly, there is whole new world beyond the paradigm of 
symmetry breaking and long range order. It is a virgin land waiting to 
be explored.

We would like to thank Pavel Etingof and Michael Freedman for useful
discussions of the mathematical aspects of topological field theory.  This
research is supported by NSF Grant No. DMR--01--23156 and by NSF-MRSEC Grant
No. DMR--02--13282.

\appendix

\section{General string-net models}
\label{genstr}

In this section, we discuss the most general string-net models. These
models can describe all doubled topological phases, including all discrete
gauge theories and doubled Chern-Simons theories. 

In these models, there is a ``spin'' degree of freedom at each branch point 
or node of a string-net, in addition to the usual string-net 
degrees of freedom. The dimension of this ``spin'' Hilbert space
depends on the string types of the $3$ strings incident on the node. 

To specify a particular model one needs to provide a $3$ index tensor 
$\del_{ijk}$ which gives the dimension of the spin Hilbert space associated 
with $\{i,j,k\}$ (in addition to the usual information). The string-net models 
discussed above correspond to the special case where $\del_{ijk} = 0,1$ for all
$i,j,k$. In the case of gauge theory, $\del_{ijk}$ is the number of 
copies of the trivial representation that appear in the tensor product 
$i \otimes j \otimes k$. Thus we need the more general string-net picture to 
describe gauge theories where the trivial representation appears multiple times
in $i \otimes j \otimes k$.

The Hilbert space of the string-net model is defined in the natural way:
the states in the string-net Hilbert space are linear superpositions of
different spatial configurations of string-nets with different spin states
at the nodes. 

One can analyze string-net condensed phases as before. The universal
properties of each phase are captured by a fixed-point ground state wave
function $\Phi$. The wave function $\Phi$ is specified by the local rules
(\ref{topinv}), (\ref{clsdst}) and simple modifications of (\ref{bubble}),
(\ref{fusion}):
\begin{align*}
 \Phi
\bpm \includegraphics[height=0.3in]{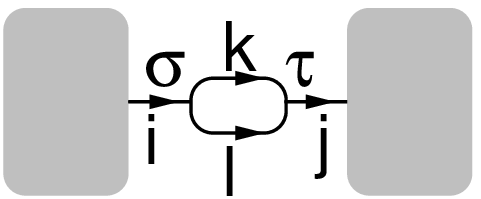} \epm  =&
\delta_{ij}\delta_{\si\tau}
\Phi 
\bpm \includegraphics[height=0.3in]{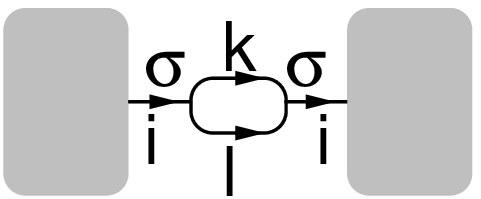} \epm
\\
 \Phi
\bpm \includegraphics[height=0.36in]{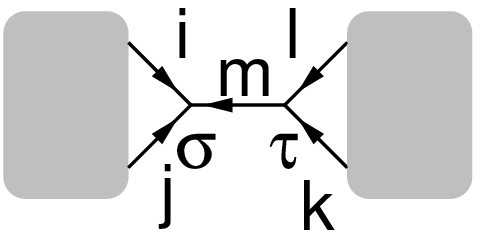} \epm  =&
\sum_{n\mu\nu} 
(F^{ijm}_{kln})^{\si\tau}_{\mu\nu}
\Phi 
\bpm \includegraphics[height=0.3in]{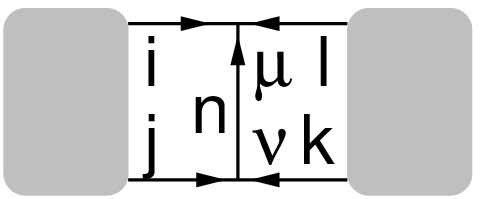} \epm
\end{align*}
The complex numerical constant $F^{ijm}_{kln}$ is now a complex tensor
$(F^{ijm}_{kln})^{\si\tau}_{\mu\nu}$ of dimension
$\del_{ijm}\times\del_{klm^*}\times\del_{inl}\times\del_{jkn^*}$. 

One can proceed as before, with self-consistency conditions, fixed-point 
Hamiltonians, string operators, and the generalization to $(3+1)$ dimensions. 
The exactly soluble models are similar to (\ref{HPi}). The main difference is
the existence of an additional spin degree of freedom at each site of the
honeycomb lattice. These spins account for the degrees of freedom at the nodes
of the string-nets.
 
\section{Self-consistency conditions}
\label{scs}

\begin{figure}[tb]
\centerline{
\includegraphics[width=2.5in]{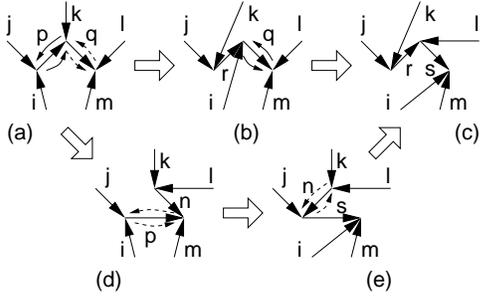}
}
\caption{
The fusion rule (\ref{fusion}) can be used to relate the amplitude of (a) to
the amplitude of (c) in two different ways. On the one hand, we can apply
the fusion rule (\ref{fusion}) twice - along the links denoted by solid
arrows - to relate (a) $\to$ (b) $\to$ (c). But we can also apply 
(\ref{fusion}) three times - along the links denoted by dashed arrows - to 
relate (a) $\to$ (d) $\to$ (e) $\to$ (c). Self-consistency requires that the 
two sequences of the operation lead to the same linear relations
between the amplitudes of (a) and (c).
}
\label{pentIdO}
\end{figure}

In this section, we derive the self-consistency conditions (\ref{pent}). We
begin with the last relation, the so-called ``pentagon identity'', since it is
the most fundamental. To derive this condition, we use the fusion rule
(\ref{fusion}) to relate the amplitude $\Phi \left(
\bmm\includegraphics[height=0.4in]{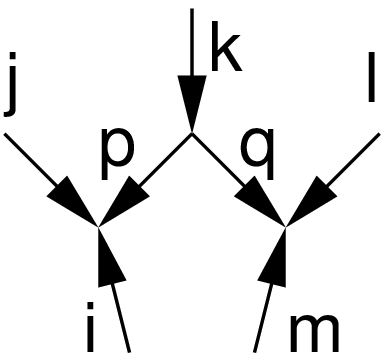}\emm \right)$ to the amplitude
$\Phi \left( \bmm\includegraphics[height=0.4in]{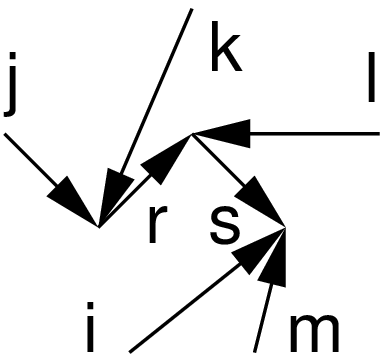}\emm \right)$ in two
distinct ways (see Fig. \ref{pentIdO}). On the one hand, we can apply the
fusion rule (\ref{fusion}) twice to obtain the relation
\begin{align*}
\Phi \left( 
\bmm\includegraphics[height=0.4in]{pen1O.eps}\emm 
\right)
=&\sum_r F^{jip}_{q^*kr^*}
\Phi \left( 
\bmm\includegraphics[height=0.4in]{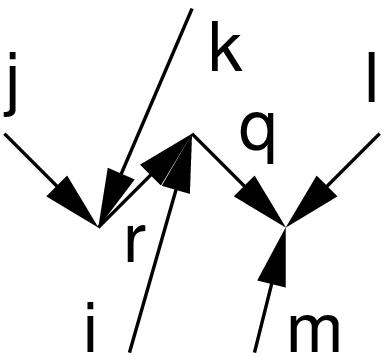}\emm
\right)  \nonumber\\
=& \sum_{r,s} F^{jip}_{q^*kr^*} F^{riq^*}_{mls^*}
\Phi \left( 
\bmm\includegraphics[height=0.4in]{pen3O.eps}\emm
\right)
\end{align*}
(Here, we neglected to draw a shaded region surrounding the whole diagram.
Just as in the local rules (\ref{topinv}-\ref{fusion}) the ends of the strings
$i,j,k,l,m$ are connected to some arbitrary string-net configuration). 
But we can also apply the fusion rule (\ref{fusion}) three times to obtain
a different relation:
\begin{align*}
\Phi \left( 
\bmm\includegraphics[height=0.4in]{pen1O.eps}\emm 
\right)
=&\sum_n F^{mlq}_{kp^*n}
\Phi \left( 
\bmm\includegraphics[height=0.4in]{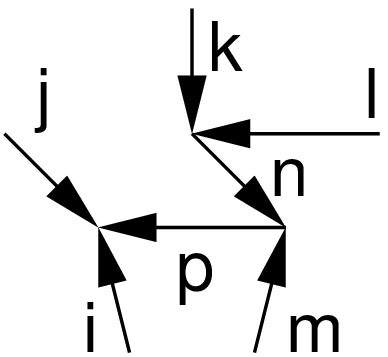}\emm
\right)  \nonumber\\
=& \sum_{n,s} F^{mlq}_{kp^*n} F^{jip}_{mns^*}
\Phi \left( 
\bmm\includegraphics[height=0.4in]{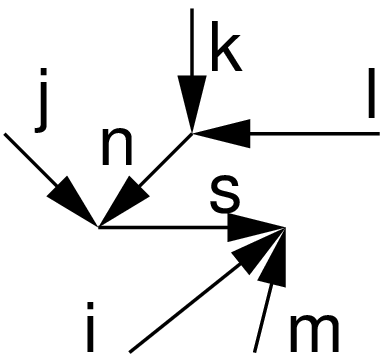}\emm
\right)
\nonumber\\
=& \sum_{n,r,s} F^{mlq}_{kp^*n} F^{jip}_{mns^*} F^{js^*n}_{lkr^*}
\Phi \left( 
\bmm\includegraphics[height=0.4in]{pen3O.eps}\emm
\right)
\end{align*}
If the rules are self-consistent, then these two relations must agree with
each other. Thus, the two coefficients of $\Phi \left(
\bmm\includegraphics[height=0.4in]{pen3O.eps}\emm \right)$ must be the same.
This equality implies the pentagon identity (\ref{pent}).

The first two relations in (\ref{pent}) are less fundamental.
In fact, the first relation is not required by self-consistency at all; it
is simply a useful convention. To see this, consider the following
rescaling transformation on wave functions $\Phi \to \t{\Phi}$. Given a 
string-net wave function $\Phi$, we can obtain a new wave function 
$\tilde{\Phi}$ by multiplying the amplitude $\Phi(X)$ for a string-net 
configuration $X$ by an arbitrary factor $f(i,j,k)$ for each vertex 
$\{i,j,k\}$ in $X$. As long as $f(i,j,k)$ is symmetric in $i,j,k$ and 
$f(0,i,i^*) = 1$, this operation preserves the topological invariance of 
$\Phi$. The rescaled wave function $\tilde{\Phi}$ satisfies the same set of 
local rules with rescaled $F^{ijm}_{kln}$:
\begin{equation}
\label{rescale}
 F^{ijm}_{kln}\to
\t F^{ijm}_{kln}=
F^{ijm}_{kln}\frac{f(i,j,m) f(k,l,m^*)}{f(n,l,i)f(j,k,n^*)}
\end{equation}
Since $\Phi$ and $\t{\Phi}$ describe the same quantum phase, we regard $F$ and
$\t{F}$ as equivalent local rules. Thus the first relation in (\ref{pent}) is
simply a normalization convention for $F$ or $\Phi$ (except when $i,j$ or $k$
vanishes; these cases require an argument similar to the derivation of the
pentagon identity).

The second relation in (\ref{pent}) has more content. This relation can be
derived by computing the amplitude for a tetrahedral string-net configuration.
We have:
\begin{align}
&\Phi\bpm\includegraphics[width=0.4in]{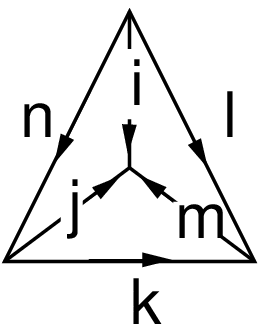}\epm
=F^{ijm}_{kln}
\Phi\bpm\includegraphics[height=0.43in]{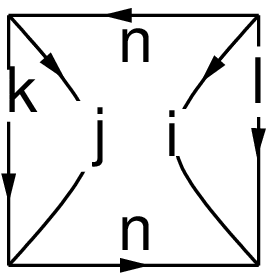}\epm
\nonumber\\
=&F^{ijm}_{kln} F^{nk^*j^*}_{kn^*0}d_k
\Phi\bpm\includegraphics[height=0.4in]{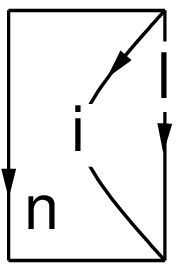}\epm
\nonumber\\
=& F^{ijm}_{kln} F^{nk^*j^*}_{kn^*0}
F^{n^*i^*l^*}_{in0} d_k d_i d_n
\Phi(\emptyset) \\
=& F^{ijm}_{kln} \w_{i}\w_{j}\w_{k}\w_{l} \Phi(\emptyset)
\end{align}

\begin{figure}[tb]
\centerline{
\includegraphics[width=3in]{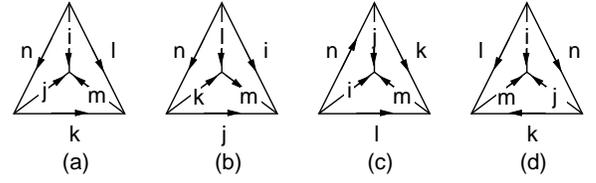}
}
\caption{
Four string-net configurations related by tetrahedral symmetry. In diagram (a),
we show the tetrahedron corresponding to $G^{ijm}_{kln}$. In diagrams (b), 
(c),(d), we show the tetrahedrons $G^{lkm^*}_{jin}$, $G^{jim}_{lkn^*}$, 
$G^{imj}_{k^*nl}$, obtained by reflecting (a) in $3$ different planes: the 
plane joining $n$ to the center of $m$, the plane joining $m$ to the center of 
$n$, and the plane joining $i$ to the center of $k$. The four tetrahedrons 
correspond to the four terms in the second relation of (\ref{pent}). 
}
\label{ijklmn2O}
\end{figure}
We define the above combination in the front of $\Phi(\emptyset)$ as:
\begin{equation}
\label{FG}
G^{ijm}_{kln}\equiv F^{ijm}_{kln} \w_{i}\w_{j}\w_{k}\w_{l} 
\end{equation} 
Imagine that the above string-net configuration lies on a sphere. In that
case, topological invariance (together with parity invariance) requires  that
$G^{ijm}_{kln}$ be invariant under all $24$ symmetries of a regular
tetrahedron. The second relation in (\ref{pent}) is simply a statement of this
tetrahedral symmetry requirement - written in terms of $F^{ijm}_{kln}$.  (See
Fig. \ref{ijklmn2O}).

In this section, we have shown that the relations (\ref{pent}) are necessary
for self-consistency. It turns out that these relations are also sufficient.
One way of proving this is to use the lattice model (\ref{HPi}). A
straightforward algebraic calculation shows that the ground state of
$\ref{HPi}$ obeys the local rules (\ref{topinv}-\ref{fusion}), as long as
(\ref{pent}) is satisfied. This establishes that the local rules are
self-consistent.

\begin{figure}[tb]
\centerline{
\includegraphics[width=2.4in]{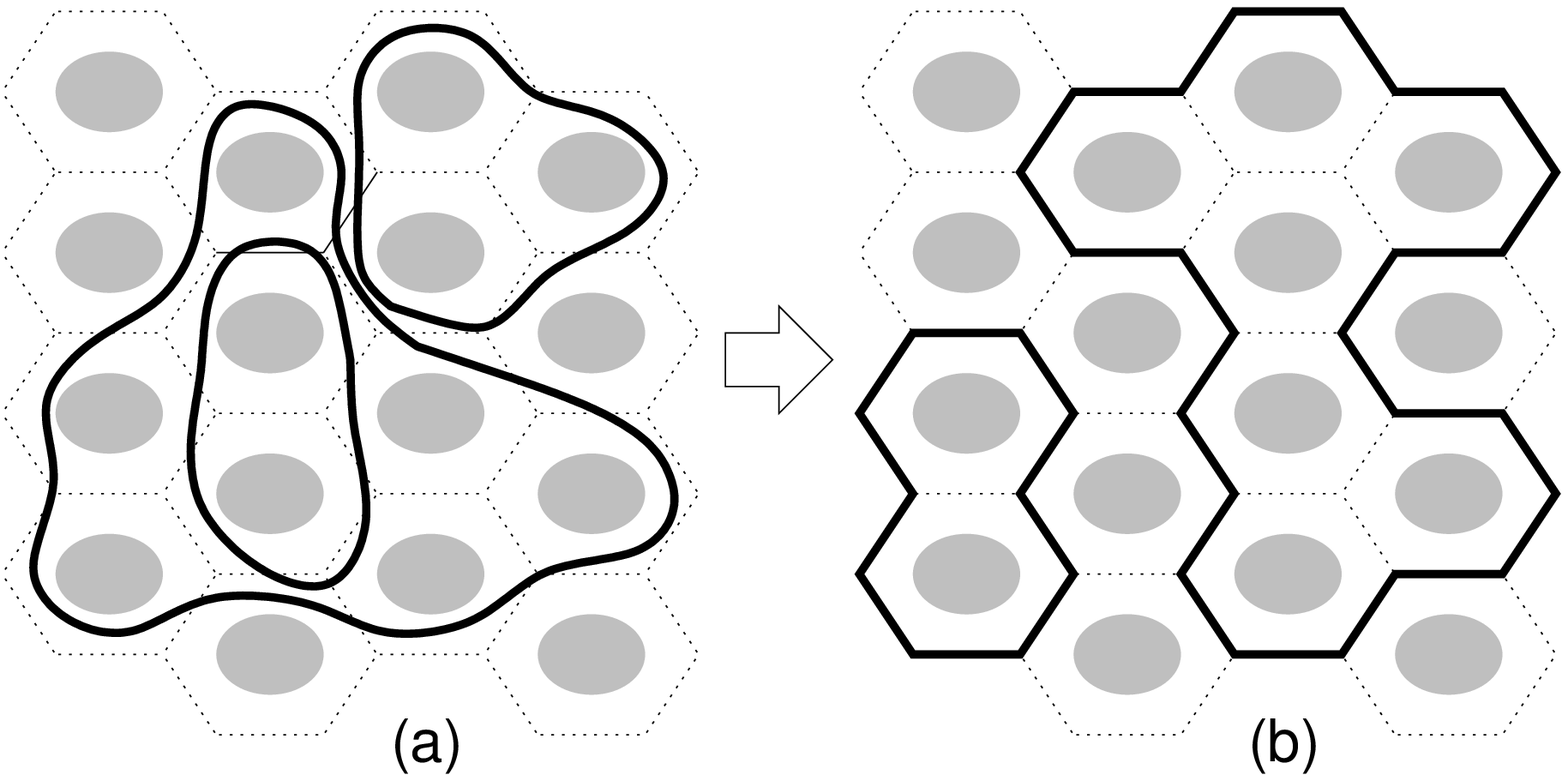}
}
\caption{
The fattened honeycomb lattice. The strings are forbidden in the shaded region.
A string state in the fattened honeycomb lattice (a) can be viewed as a
superposition of string states on the links (b).
}
\label{FHlatt}
\end{figure}

\section{Graphical representation of the Hamiltonian}
\label{grphHam}

In this section, we provide an alternative, graphical, representation of the
lattice model (\ref{HPi}). This graphical representation provides a simple
visual technique for understanding properties (a)-(c) of the Hamiltonian
(\ref{HPi}).

We begin with the $2D$ honeycomb lattice. Imagine we fatten the links of the
lattice into stripes of finite width (see Fig. \ref{FHlatt}). Then, any
string-net state in the fattened honeycomb lattice (Fig. \ref{FHlatt}a) can be
viewed as a superposition of string-net states in the original, unfattened
lattice (Fig. \ref{FHlatt}b). This mapping is obtained via the local rules
(\ref{topinv}-\ref{fusion}). Using these rules, we can relate the amplitude
$\Phi(X)$ for a string-net in the fattened lattice to a linear combination of
string-net amplitudes in the original lattice: $\Phi(X) = \sum a_i \Phi(X_i)$.
This provides a natural linear relation between the states in the fattened
lattice and those in the unfattened lattice: $|X\> = \sum a_i |X_i\>$. This
linear relation is independent of the particular way in which the local rules
(\ref{topinv}-\ref{fusion}) are applied, as long as the rules are
self-consistent. 

\begin{figure}
\centerline{
\includegraphics[width=2.4in]{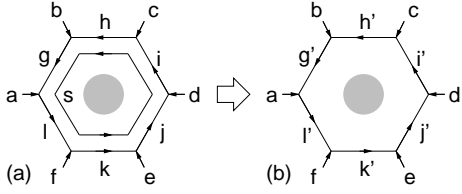}
}
\caption{
The action of $B^s_{\v p}$ is equivalent to adding a loop of 
type-$s$ string. The resulting string-net state (a) is actually a linear 
combination of the string-net states (b). The coefficients in this linear 
relation can be obtained by using the local rules (\ref{topinv}-\ref{fusion}) 
to reduce (a) to (b).  
}
\label{BsLoopO}
\end{figure}

In this way, the fattened honeycomb lattice provides an alternative notation
for representing the states in the Hilbert space of (\ref{HPi}). This notation
is useful because the magnetic energy operators $B^{s}_{\v p}$ are simple in
this representation. Indeed, the action of the operator $B^s_{\v p}$ on the
string-net state $\Big| \bmm\includegraphics[height=0.6in]{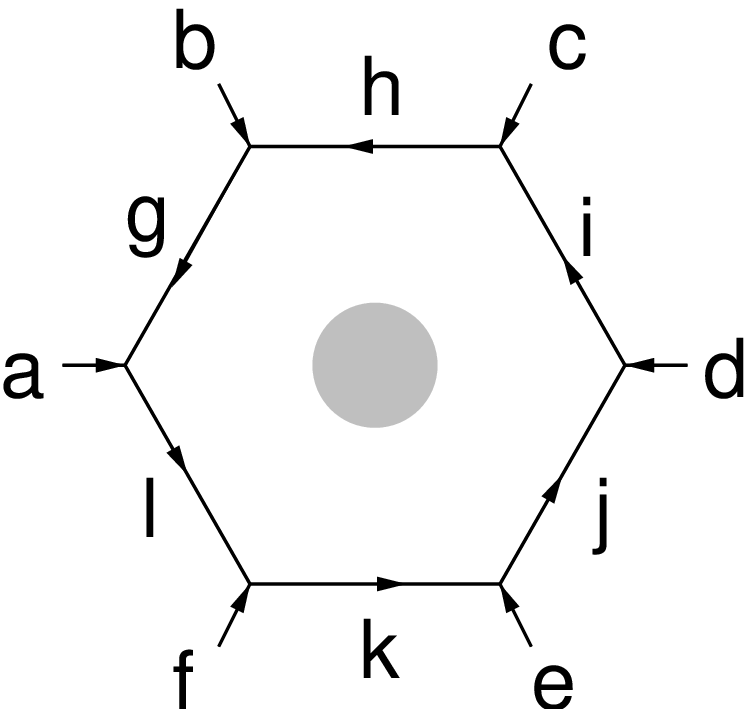}\emm \Big
\>$ is equivalent to simply adding a loop of type-$s$ string:
\begin{align*}
&
B^s_{\v p}
\Big|
\bmm\includegraphics[height=0.8in]{BsL0O.eps}\emm 
\Big \>
=
\Big|
\bmm\includegraphics[height=0.8in]{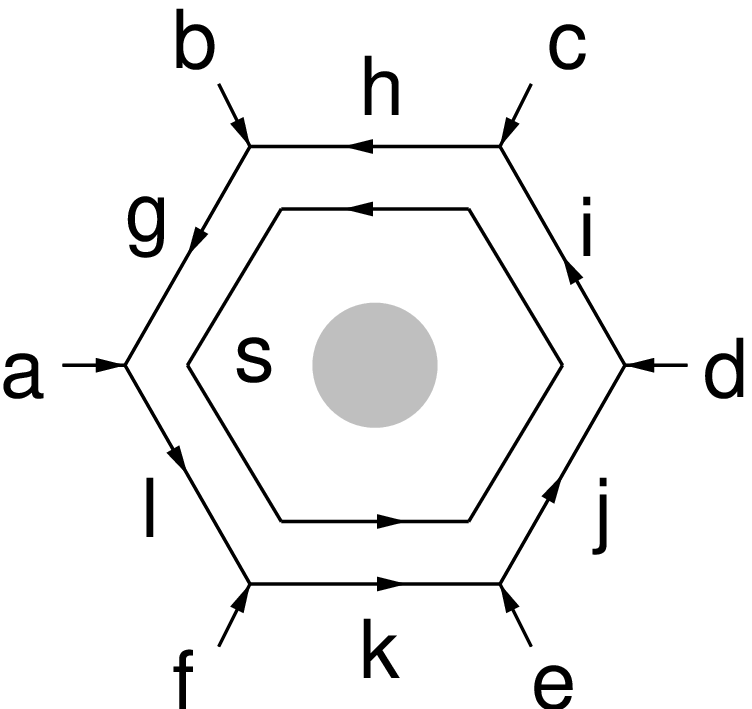}\emm 
\Big\>  
\end{align*}
As we described above, we can use the local rules (\ref{topinv}-\ref{fusion}) 
to rewrite $\Big| \bmm\includegraphics[height=0.6in]{BsL1O.eps}\emm \Big \>$ as
a linear combination of the physical string-net states with strings only on
the links, that is to reduce Fig. \ref{BsLoopO}a to Fig. \ref{BsLoopO}b.  This
allows us to obtain the matrix elements of $B^s_{\v p}$. 

The following is a particular way to implement the above procedure:
\begin{widetext}
\begin{align}
\label{B6G}
&
B^s_{\v p}
\Big| 
\bmm\includegraphics[height=0.8in]{BsL0O.eps}\emm 
\Big\> 
=
\Big| 
\bmm\includegraphics[height=0.8in]{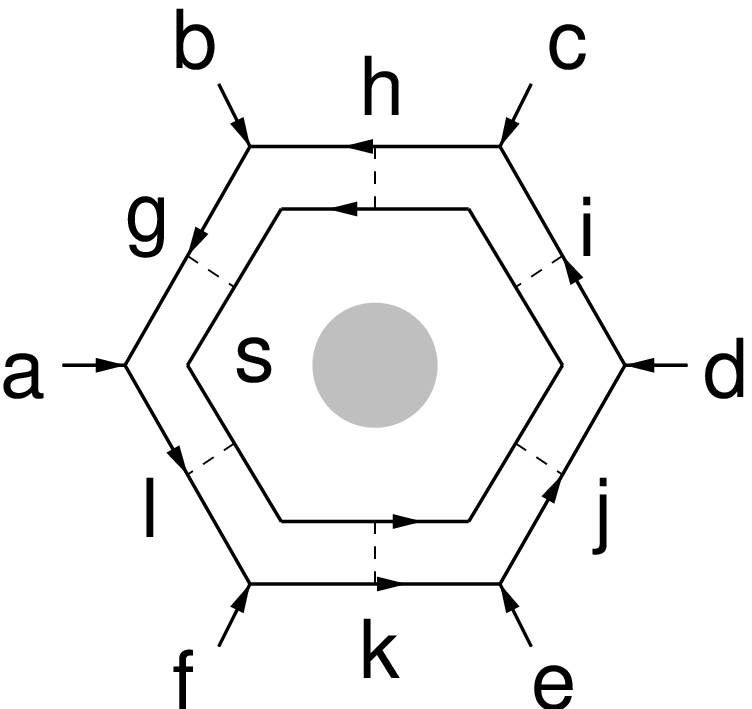}\emm 
\Big\> 
= \sum_{g'h'i'j'k'l'} 
F^{gg^*0}_{s^*sg'^*}
F^{hh^*0}_{s^*sh'^*}
F^{ii^*0}_{s^*si'^*}
F^{jj^*0}_{s^*sj'^*}
F^{kk^*0}_{s^*sk'^*}
F^{ll^*0}_{s^*sl'^*}
\Big| 
\bmm\includegraphics[height=0.8in]{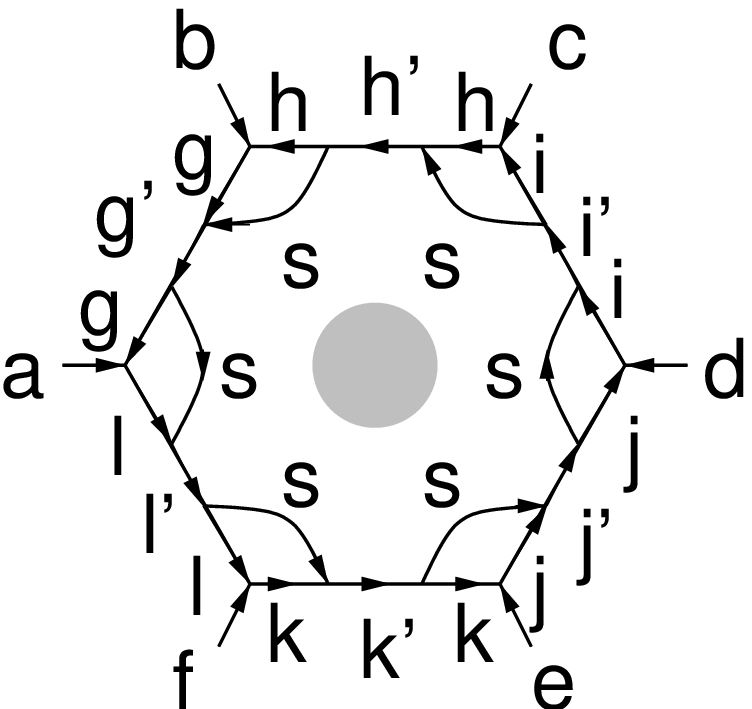}\emm 
\Big\>
\nonumber\\
=& 
\sum_{g'h'i'j'k'l'} 
F^{gg^*0}_{s^*sg'^*}
F^{hh^*0}_{s^*sh'^*}
F^{ii^*0}_{s^*si'^*}
F^{jj^*0}_{s^*sj'^*}
F^{kk^*0}_{s^*sk'^*}
F^{ll^*0}_{s^*sl'^*}
F^{bg^*h}_{s^*h'g'^*}
F^{ch^*i}_{s^*i'h'^*}
F^{di^*j}_{s^*j'i'^*}
F^{ej^*k}_{s^*k'j'^*}
F^{fk^*l}_{s^*l'k'^*}
F^{al^*g}_{s^*g'l'^*}
\Big| 
\bmm\includegraphics[height=0.8in]{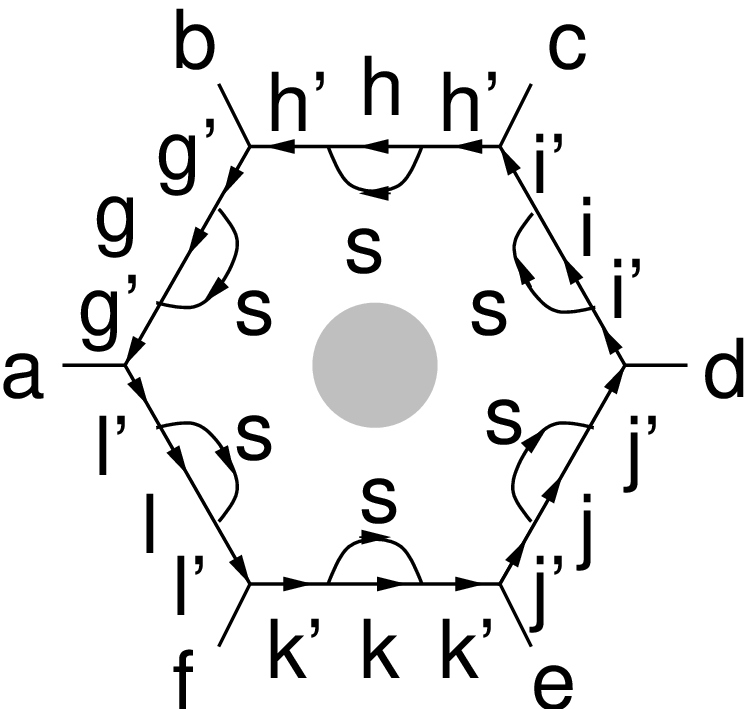}\emm 
\Big\>  
\nonumber\\
=&
\sum_{g'h'i'j'k'l'} 
F^{bg^*h}_{s^*h'g'^*}
F^{ch^*i}_{s^*i'h'^*}
F^{di^*j}_{s^*j'i'^*}
F^{ej^*k}_{s^*k'j'^*}
F^{fk^*l}_{s^*l'k'^*}
F^{al^*g}_{s^*g'l'^*}\Big|
\bmm\includegraphics[height=0.8in]{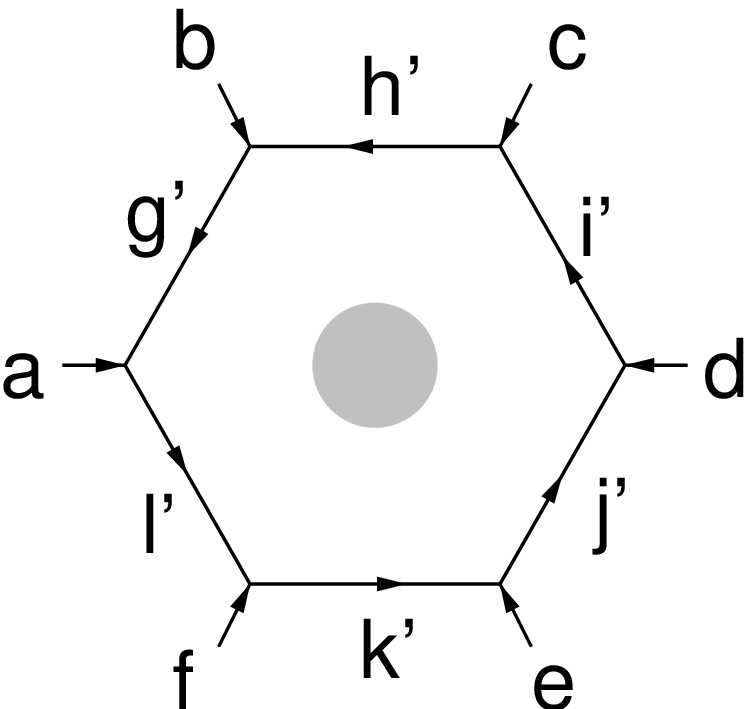}\emm
\Big\>
\end{align}
\end{widetext}
Notice that \Eq{B6G} is exactly \Eq{B6F}. Thus, the graphical representation
of $B^{s}_{\v p}$ agrees with the original algebraic definition.

\begin{figure}[tb]
\centerline{
\includegraphics[width=3.5in]{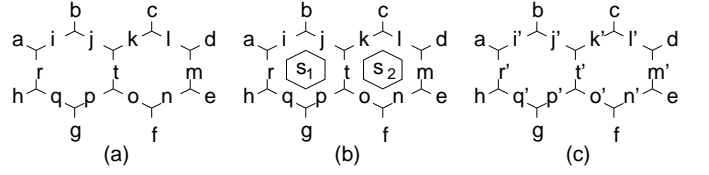}
}
\caption{
The action of $B^{s_1}_{\v{p_1}}B^{s_2}_{\v{p_2}}$ on the string-net state (a)
can be represented by adding two loops of type-$s_1$ and type-$s_2$ strings
as shown in (b). The string-net state (b) is a linear combination of the 
string-net states (c). The coefficients are obtained by using (\ref{topinv}
-\ref{fusion}) to reduce (b) to (c).
}
\label{BBcom}
\end{figure}

Using the graphical representation of $B^s_{\v p}$ we can easily show that
$B^{s_1}_{\v{p_1}}$ and $B^{s_2}_{\v{p_2}}$ commute. The derivation is much
simpler then the more straightforward algebraic calculation. First note that
these operators will commute if $\v {p_1}$, $\v {p_2}$ are well-separated.
Thus, we only have to consider the case where $\v{p_1}$ and $\v{p_2}$ are
adjacent, or the case where $\v{p_1}$, $\v{p_2}$ coincide. We begin with the
nearest neighbor case. The action of $B^{s_1}_{\v{p_1}}B^{s_2}_{\v{p_2}}$ on
the string-net state Fig. \ref{BBcom}a can be represented as Fig.
\ref{BBcom}b.  Fig.  \ref{BBcom}b can then be related to a linear combination
of the string-net states shown in Fig.  \ref{BBcom}c. The coefficients in this
relation are the matrix elements of $B^{s_1}_{\v p_1}B^{s_2}_{\v p_2}$.  But
by the same argument, the action of $B^{s_2}_{\v p_2}B^{s_1}_{\v p_1}$ can
also be represented by Fig. \ref{BBcom}b. We conclude that $B^{s_2}_{\v
p_2}B^{s_1}_{\v p_1}$, $B^{s_1}_{\v p_1}B^{s_2}_{\v p_2}$ have the same matrix
elements. Thus, the two operators commute in this case. 

On the other hand, when $\v p_1=\v p_2$, we have
\begin{align}
&
B^{s_2}_{\v p} B^{s_1}_{\v p}
\Big| 
\bmm\includegraphics[height=0.6in]{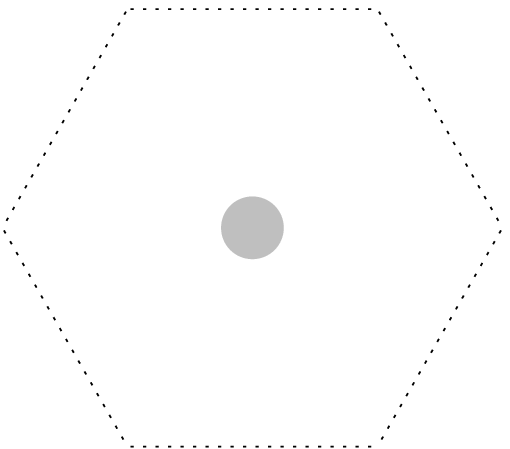}\emm 
\Big\>
=
\Big| 
\bmm\includegraphics[height=0.6in]{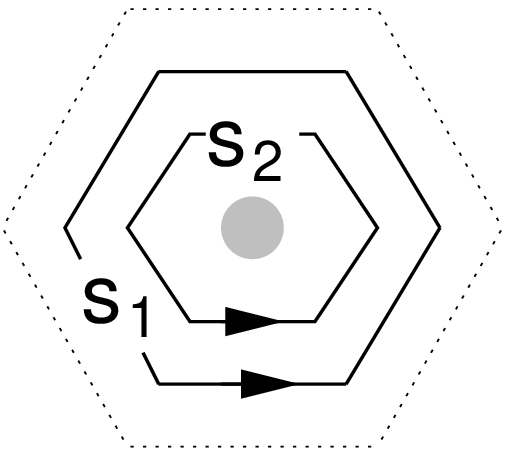}\emm 
\Big\>
\nonumber \\
=&
\sum_k 
F^{s_1s_1^*0}_{s_2^*s_2k^*}
\Big| 
\bmm\includegraphics[height=0.6in]{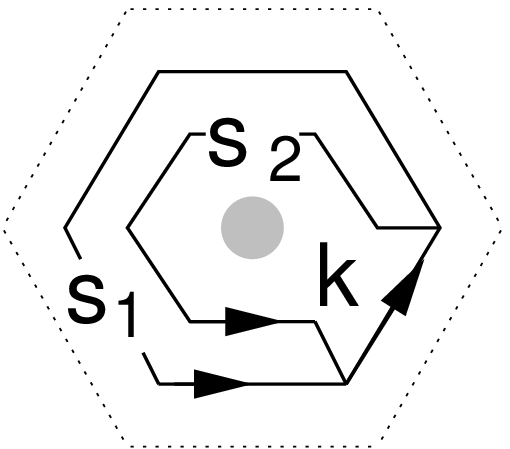}\emm 
\Big\>
\nonumber\\
=&
\sum_k 
F^{s_1s_1^*0}_{s_2^*s_2k^*}F^{k^*s_2s_1}_{s_2^*k0}d_{s_2^*}
\Big| 
\bmm\includegraphics[height=0.6in]{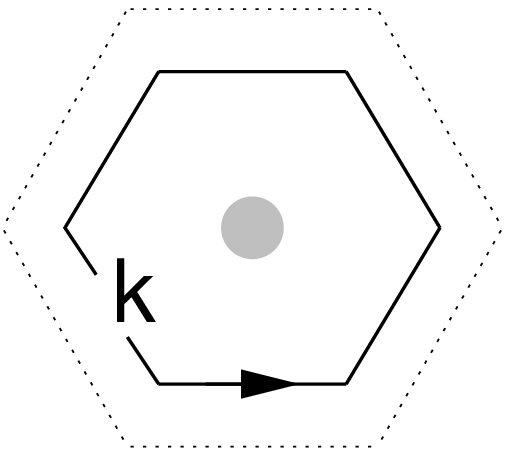}\emm 
\Big\>
\nonumber \\
=&
\sum_k
\del_{k^*s_2s_1} \Big| 
\bmm\includegraphics[height=0.6in]{BBc3O.eps}\emm 
\Big\>
\end{align}
Thus,
\begin{equation}
\label{BBB}
B^{s_2}_{\v p} B^{s_1}_{\v p}= 
\sum_k \del_{k^*s_2s_1}  B^k_{\v p} .
\end{equation}
Since $\del_{k^*s_2s_1}$ is symmetric in $s_2$, $s_1$, we conclude that
$B^{s_1}_{\v p} B^{s_2}_{\v p}=B^{s_2}_{\v p} B^{s_1}_{\v p}$, so the
operators commute in this case as well. This establishes property (a) 
of the Hamiltonian (\ref{HPi}).

Equation \Eq{BBB} also sheds light on the spectrum of the $B^s_{\v p}$ 
operators. Let the simultaneous eigenvalues of $B^s_{\v p}$ (with $\v p$ fixed)
be $\{b^{s}_{q}\}$. Then, by \Eq{BBB} these eigenvalues satisfy
\begin{displaymath}
\sum_k \del_{k^*s_2s_1} b^k_q =b^{s_2}_{q} b^{s_1}_{q}
\end{displaymath}
We can view this as an eigenvalue equation for the $(N+1) \times (N+1)$ 
matrix $M_{s_2}$, defined by $M^{i}_{s_2,j} = \del_{j^*s_2i}$. The
simultaneous eigenvalues $b^{s_2}_{q}$ are simply the simultaneous
eigenvalues of the matrices $M_{s_2}$. In particular, this means that the index
$q$ ranges over a set of size $N+1$.

Each value of $q$ corresponds to a different possible state for the plaquette 
$\v p$. The magnetic energies of these $N+1$ different states are given by:
$E_q = -\sum_{s} a_s b^s_q$. Depending on the parameter choice $a_s$, all
on the plaquettes $\v p$ will be in one of these states $q$. In this way, the 
Hamiltonian (\ref{HPi}) can be in $N+1$ different quantum phase. This 
establishes property (b) of the Hamiltonian (\ref{HPi}).

One particular state $q$ is particularly interesting. This state corresponds
to the simultaneous eigenvalues $b^{s} = d_{s}$. It is not hard to show 
that the parameter choice $a_s = \frac{d_s}{\sum_k d_k^2}$ makes this
state energetically favorable. In fact, using (\ref{BBB}) one can show that
$B_{\v p}$ is a projector for this parameter choice, and that $B_{\v p} = 1$
for this state. 

Furthermore, the ground state wave function for this parameter choice obeys the
local rules (\ref{topinv}-\ref{fusion}). One way to see this is to compare 
$B_{\v p} \Big|\bmm\includegraphics[width=0.45in]{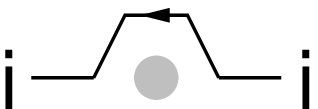}\emm \Big\>$ with 
$B_{\v p} \Big|\bmm\includegraphics[width=0.45in]{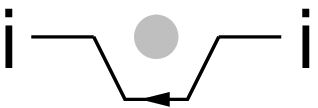}\emm \Big\>$.
For the first state, we find
\begin{align*}
B_{\v p}
\Big|
\bmm\includegraphics[width=0.45in]{hop0O.eps}\emm 
\Big\> 
=&\sum_s a_s
\Big|
\bmm\includegraphics[width=0.45in]{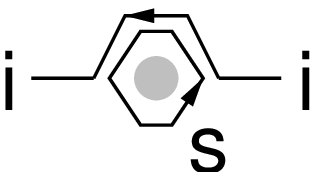}\emm 
\Big\> \\
=&\sum_{j,s} \frac{d_s}{\sum_k d_k^2} F^{ii^*0}_{s^*sj^*}
\Big|
\bmm\includegraphics[width=0.45in]{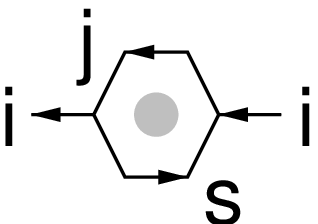}\emm 
\Big\> \\
=&\sum_{j,s} \frac{\w_j\w_s}{\w_i \sum_k d_k^2}
\Big|
\bmm\includegraphics[width=0.45in]{hop3O.eps}\emm 
\Big\>
\end{align*}
For the second state, we find the same result:
\begin{align*}
&
B_{\v p}
\Big|
\bmm\includegraphics[width=0.45in]{hop0aO.eps}\emm 
\Big\>
=\sum_{j,s} \frac{\w_j\w_s}{\w_i \sum_k d_k^2}
\Big|
\bmm\includegraphics[width=0.45in]{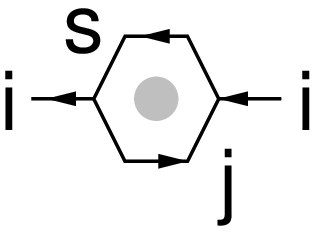}\emm 
\Big\>
\end{align*}
It follows that
\begin{align*}
0 &= \Big\<\bmm\includegraphics[width=0.45in]{hop0O.eps}\emm \Big| B_{\v p} 
\Big|\Phi \Big\> -
\Big\<\bmm\includegraphics[width=0.45in]{hop0aO.eps}\emm \Big| B_{\v p} \Big|
\Phi \Big\> \\
&=
\Big\<\bmm\includegraphics[width=0.45in]{hop0O.eps}\emm \Big| \Phi \Big\> -
\Big\<\bmm\includegraphics[width=0.45in]{hop0aO.eps}\emm \Big| \Phi \Big\> \\
\end{align*}
so
\begin{displaymath}
\Phi 
\bpm\includegraphics[width=0.45in]{hop0O.eps}\epm
=
\Phi 
\bpm\includegraphics[width=0.45in]{hop0aO.eps}\epm
\end{displaymath}
This result means that the strings can be moved through the forbidden
regions at the center of the hexagons. Thus, the local rules which
were originally restricted to the fattened honeycomb lattice can be extended
throughout the entire $2D$ plane. The wave function $\Phi$ obeys these
continuum local rules and has a smooth continuum limit.
We call such a state smooth topological state.
This establishes property (c) of the Hamiltonian
(\ref{HPi}). 

The wave functions of some smooth topological states are positive definite.
So those wave functions can be viewed as the statistical weights of certain
statistical models in the same spatial dimensions.
What is interesting is that those statistical models are local models
with short-ranged interactions \cite{W8929,W9085,AFF0493}.

\section{Graphical representation of the string operators}
\label{grphStr}
\begin{figure}[tb]
\centerline{
\includegraphics[width=2.5in]{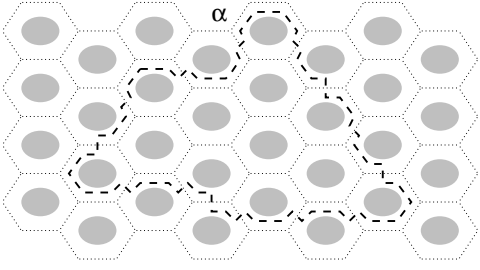}
}
\caption{
The action of the string operator $W_{\al}(P)$ is equivalent to adding a 
type-$\al$ string along the path $P$. The resulting string-net state
can be reduced to a linear combination of states on the honeycomb lattice,
using the local rules (\ref{topinv}-\ref{fusion}), (\ref{brdrl}).
}
\label{LngStr}
\end{figure}

In this section, we describe a graphical representation of the long string
operators $W_{\al}(P)$. Just as in the previous section, this representation
involves the fattened honeycomb lattice. The action of the string operator
$W_{\al}(P)$ on a general string state $X$, is simply to create a string
labeled $\al$ along the path $P$. The resulting string-net state can then be
reduced to a linear combination of string-net states on the unfattened
lattice. The coefficients in this linear combination are the matrix elements
of $W_{\al}(P)$.

However, none of the rules (\ref{topinv}-\ref{fusion}) involve strings labeled
$\al$, nor do they allow for crossings. Thus, the reduction to string-net
states on the unfattened lattice requires new local rules. These new local
rules are defined by the $4$ index objects $\Om^{j}_{\al,sti}$,
$\bar{\Om}^{j}_{\al,sti}$, and the integers $n_{\al,i}$:
\begin{eqnarray}
 \Big| \bmm\includegraphics[height=0.3in]{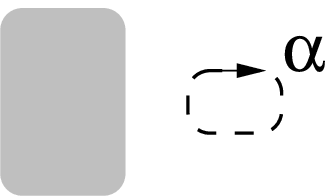}\emm \Big\>
&=& \sum_{i} n_{\al,i} 
\Big| \bmm\includegraphics[height=0.3in]{Xil0.eps}\emm \Big\> \nonumber \\
 \Big| \bmm\includegraphics[height=0.3in]{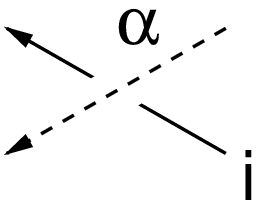}\emm \Big\>
&=& \sum_{jst}
(\Om^{j}_{\al,sti})_{\si\tau}
 \Big| \bmm\includegraphics[height=0.3in]{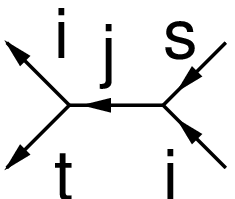}\emm \Big\> \nonumber \\
 \Big| \bmm\includegraphics[height=0.3in]{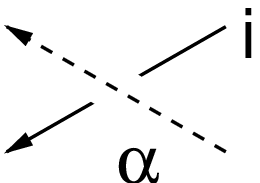}\emm \Big\>
&=& \sum_{jst}
(\bar{\Om}^{j}_{\al,sti})_{\si\tau}
 \Big| \bmm\includegraphics[height=0.3in]{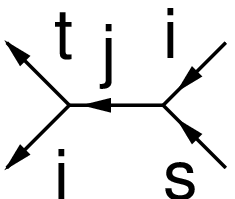}\emm \Big\> 
\label{brdrl}
\end{eqnarray}
Here, $\si,\tau$ are the two indices of the matrix $\Om^{ij}_{st}$. (Until
now, we've neglected to write out these indices explicitly). 

After applying these rules, we then need to join together the resulting
string-nets. The ``joining rule'' for two string types $s_1$, $s_2$ is as
follows. If $s_1 \neq s_2$, we don't join the two strings: we simply throw
away the diagram. If $s_1 = s_2$, then we join the two strings and contract
the two corresponding indices $\si_1$, $\si_2$. That is, we multiply the two
$\Omega$ matrices together in the usual way. Using the same approach as
\Eq{B6G}, one can show that the graphical definition of $W_{\al}(P)$ agrees
with the algebraic definition (\ref{strop}).

In the previous section, we used the graphical representation of $B^s_{\v p}$
to show that these operators commute. The string operators $W_{\al}(P)$ can be
analyzed in the same way. With a simple graphical argument one can show that
the string operators $W_{\al}(P)$ commute with the magnetic operators $B^s_{\v
p}$ provided that (\ref{topinv}-\ref{fusion}),(\ref{brdrl}) satisfy the
conditions
\begin{eqnarray}
\Big| \bmm\includegraphics[width=0.45in]{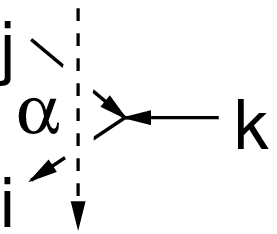}\emm\Big\>
&=& \Big| \bmm\includegraphics[width=0.45in]{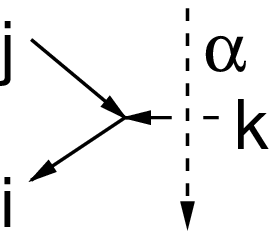}\emm \Big\> \\
 \Big| \bmm\includegraphics[height=0.3in]{Brd1aO.eps}\emm \Big\>
&=& \Big| \bmm\includegraphics[height=0.22in]{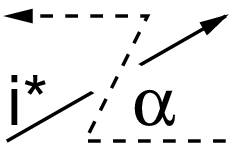}\emm \Big\> 
\end{eqnarray}
These relations are precisely the commutativity conditions (\ref{omeq}),
written in graphical form.

\bibliography{/home/wen/bib/wencross,/home/wen/bib/all,/home/wen/bib/publst} 

\end{document}